# Drastic modification in thermal conductivity of TiCoSb Half-Heusler alloy: Phonon engineering by lattice softening and ionic polarization

Suman Mahakal[1], Avijit Jana[1], Diptasikha Das[2], Nabakumar Rana[3], Pallabi Sardar[2], Shamima Hussain[4], Santanu K. Maiti[5], and Kartick Malik*[1]

[1] *Department of Physics, Vidyasagar Metropolitan College, Kolkata-700 006, India.*
[2] *Department of Physics, Adamas University, Kolkata-700 126, India.*
[3] *Department of Physics, University of Calcutta, Kolkata-700 009, India.*
[4] *UGC-DAE Consortium for Scientific Research, Kalpakkam Node, Kokilamedu-603 104, Tamil Nadu, India.*
[5] *Physics and Applied Mathematics Unit, Indian Statistical Institute, 203 Barrackpore Trunk Road, Kolkata-700 128, India.*

**Abstract**

A drastic variation (~47%) in thermal conductivity (κ) for synthesized samples ($TiCoSb_{1+x}$, x=0.0, 0.01, 0.02, 0.03, 0.04, and 0.06) is observed. The lowest κ is reported for the $TiCoSb_{1.02}$ sample. Thermal variation of κ is estimated from the temperature and power-dependent Raman spectroscopy data. Embedded phases and Co vacancy are analysed, employing scanning electron microscopy and transmission electron microscopy data. X-ray absorption fine structure (XAFS) spectroscopy reveals the Co vacancy in synthesized samples, and the most ordered phase is $TiCoSb_{1.02}$ amid the synthesized samples. X-ray photoelectron spectroscopy measurement of the synthesized samples provides direct evidence of Co vacancies and their increase with Sb concentration (x). Lattice dynamics are revealed using Raman Spectroscopy (RS) measurements. RS data accomplishes that variation in κ as a function of Sb concentration is observed owing to an alteration in phonon group velocity, related to lattice softening. The polar nature of the TiCoSb half-Heusler (HH) sample is revealed. Longitudinal Optical and Transvers Optical phonon (LO-TO) splitting in RS is observed due to the polar nature of $TiCoSb_{1+x}$ synthesized samples. Tailoring in LO-TO splitting due to the screening effect, correlated with Co vacancies, is reported for $TiCoSb_{1+x}$ synthesized samples. Lattice softening and LO-TO splitting lead to a minimum κ for the $TiCoSb_{1.02}$ synthesized sample.

## I. Introduction

Thermoelectric (TE) materials have the potential to recover waste heat, generated during energy conversion. Investigation of TE technology and related materials is one of the interesting topics for researchers due to alternative energy sources, an intriguing class of materials, and unique physics involved in the technology.[1] TE materials are those, which convert thermal energy to electrical energy and vice versa.[1] However, interrelated materials properties, involved in the dimensionless figure of merit ($ZT = S^2\sigma T/\kappa$, where S is thermopower, σ is electrical conductivity, κ is the total thermal conductivity, and T is absolute temperature), limit the material development. Recently, research is focused on enhancing ZT by decreasing κ ($\kappa = \kappa_e + \kappa_l$, $\kappa_e$ is the electronic part and $\kappa_l$ is the lattice part) of TE materials. $\kappa_l$ is the only decoupled parameter amid the properties involved in ZT.[2] Several reports indicate that high Power Factor (PF), $S^2\sigma$ may lead to enhancement of ZT.[2] However, commercial implementation is limited due to high κ.

MCoSb and MNiSn (M=Ti, Hf and Zr) based half-Heusler (HH) compounds are promising candidates for TE technology.[3] This is attributed to their narrow band gap, high substitutability of constituent elements, enhanced contact engineering, and high thermal stability.[4] HH alloys are in the category of XYZ type compound, where X is early transition metal (Sc, Ti, V groups), Y is late transition metal (Fe, Co, Ni groups), i.e., less electropositive than X, and Z is main group elements.[5] Preliminary research on the HH materials commenced in the late 1990s, identifying n-type and p-type phases with $X^{IV}NiSn$ and $X^{IV}CoSb$ ($X^{IV}$ = Ti, Zr, and Hf), respectively.[6] Till now, various HH-based TE materials such as TiCoSb, TiNiSn, $X^{V}FeSb$ ($X^{V}$ = V, Nb, and Ta), NbCoSn, ZrCoBi, etc.[7] are well investigated, exhibiting high PF. TiCoSb is one of the potential p-type TE materials[3] and crystallized as MgAgAs type structure with atomic positions at 4a (0, 0, 0), 4b (1/2, 1/2, 1/2), 4c (1/4, 1/4, 1/4).[8] Narrow band gap with Fermi level on top of the valence band (p-type) and 18 valence electron count (VEC) have made them promising candidates in TE power generation.

TiCoSb alloy draws the attention of the researchers as a TE material due to substantial S and moderate σ. However, concurrent exhibition of high κ of TiCoSb and other HH TE materials leads to a reduction in the overall ZT.[4] Resurgence is focused on enhancing the efficiency by reducing κ for TiCoSb.[9] Nano inclusion and complex iso-electronic alloying are important routes to enhance the efficiency of a TE material, by reducing κ.[9] Micro/nanostructuring of materials also leads to a decrease in κ.[10] Huang et al. have reported that incorporation of $ZrO_2$ nanoparticles in the ZrNiSn HH alloy reduces κ significantly (~35%) through the introduction of phonon scattering centers.[11] Composite material, 2% $Zr_{0.25}Hf_{0.75}Ni_2Sn$ Full-Heusler (FH) phases into the $Zr_{0.25}Hf_{0.75}NiSn$ HH matrix causes a reduction in κ (~20%).[12] Further, the introduction of point defects and disorder may lead to a substantial reduction in κ through enhanced phonon scattering. The effect of iso-electronic alloying, band engineering, and nanostructuring on TE properties is extensively investigated.[9] However, it is crucial to mention that $\kappa_l$ is highly responsive to the variation in phonon dispersion, specifically the speed of phonon in material.[13] Tan et al. have reported a dramatic reduction of κ in SnTe and $AgSbTe_2$ alloy, owing to a reduced velocity of phonons.[14] Hanus et al. have highlighted that the inclusion of internal-strain-induced lattice softening proves to be a promising approach to reduce κ.[13] S. Y. Back et al. have also observed 25% reduced $\kappa_l$ in InTe TE material due to lattice softening.[15] Hence, lattice softening is one of the important approaches to optimize $\kappa_l$.[14, 15] Polar optical phonon scattering in polar TE materials also plays a very important role in electronic and thermal transport.[16] It should be noted that the polar coupling constant for various HH TE materials (ZrCoSb~0.89, NbFeSb~0.74, ZrNiSn~0.32 etc) shows higher value than other well-known PbTe based (PbTe~0.29, PbSe~0.36) TE materials.[17] Presence of defects in polar TE materials has the potential to alter the scattering centers, may influence both σ and κ.[18] In a recent study, Ren et al. have reported anomalous phononic transport in ZrNiBi HH alloys due to the presence of vacancies.[18] Substitution of Sn at Co site in ZrCoSb TE material enhances the intensity of phonon–phonon interaction owing to the dominant role of Co/4d Frenkel point defects, results a decrease in κ (~80%).[19]

There are several studies on the reduction of $\kappa_l$ through modification in structure and embedded phases by doping and dispersion.[20, 21] $\kappa_l$ is strongly influenced by the scattering of long to mid-range phonons.[22] It is crucial to note that phonon vibration and local structure are strongly corroborated. Modification in the local structure may be another approach to tune the $\kappa_l$.[23] X-ray absorption fine structure (XAFS) spectroscopy is one of the powerful tools to explore the local crystalline structure.[24] The local environment surrounding a particular atomic species within a compound or alloy is frequently investigated using the extended X-ray absorption fine structure (EXAFS) region of the XAFS spectrum.[24] In addition, the near-edge region of the spectrum, i.e., X-ray absorption near-edge structure (XANES) is considered a good stereochemical probe for remarkable sensitivity to bonding geometry. Local structure of several materials is investigated using XAFS studies in the last few years.[24] Low $\kappa_l$ owing to Ni disorder in NiZrSn HH alloy is revealed from XAFS study, reported by Miyazaki et al.[25] It is crucial to note that electronic and thermal properties of compounds may be optimized by tuning the local ordering of atoms.[25] Structural defects and disorder result in alterations of phonon dynamics and have an impact on $\kappa_l$.[26] Raman spectroscopy (RS) is a non-destructive tool to reveal the phonon dynamics of the TE materials. Phonon anharmonicity in Se-doped $Bi_2Te_3$ TE material is explored using RS.[27] Patel et al. have investigated phonon anharmonicity and optical phonon decay mechanisms in $Hf_xSn_{1-x}Se_2$ (x=0-0.2).[28] Lattice dynamics of topological insulator, an intriguing class of material $Bi_{1.9}Dy_{0.1}Te_3$ is also investigated by RS.[29] However, the obvious consequences of defect and disorder are the broadening in phonon vibrational modes and softening of the lattice.[30] Putra et al. have investigated the lattice dynamics and anharmonicity in ZnO-buffered $MgB_2$ employing RS.[31]

In this attempt, we try to investigate the nature of interactions associated with phonon transport and reveal the effect of Sb concentration on κ for $TiCoSb_{1+x}$ ($0.0 \leq x \leq 0.06$) samples. A detailed experimental study of the lattice dynamics of the TiCoSb HH alloy is explored. The role of defects as well as embedded phases on κ and crystalline order of polycrystalline TiCoSb alloys is investigated in this article. Presence of Co vacancy, and embedded phases are revealed from X-ray diffraction (XRD), scanning electron microscopy (SEM), energy-dispersive X-ray spectroscopy (EDS), Transmission electron microscopy (TEM), X-ray absorption fine structure (XAFS) spectroscopy, and X-ray photoelectron spectroscopy (XPS). XAFS study reveals that the crystalline order of the synthesized TiCoSb-based HH TE materials is mainly preserved owing to the contribution of the 2nd shell, i.e., Co-Sb shell. Raman spectroscopy measurement is employed for an in-depth study of lattice dynamics and vibrational spectroscopy of the synthesized TiCoSb HH alloy. The polar nature of TiCoSb is observed, and a plausible explanation is provided from experimental data. Lattice softening, splitting of longitudinal optical and transverse optical phonon (LO-TO splitting), and the screening effect play a crucial role in determining the value of κ in synthesized samples. We have specifically emphasized to elucidate the low κ of the synthesized $TiCoSb_{1+x}$ HH alloys and, correlation with structural, stereochemical, and vibrational properties.

The key findings of our study are (i) Co vacancies and embedded phases in $TiCoSb_{1+x}$ unitcell increase with Sb concentration, supported by EDS, TEM, XAFS and XPS, (ii) most ordered TiCoSb sample amid the synthesized $TiCoSb_{1+x}$ is $TiCoSb_{1.02}$, revealed from XAFS as well as RS measurements, (iii) polar nature of $TiCoSb_{1+x}$ alloys are supported from the LO-TO splitting in RS, and (iv) the minimum value of κ (κ~3.9 W/mK), as per our knowledge is achieved for pristine TiCoSb sample, tuning crystalline strain and Co vacancies.

The rest part of the work is arranged as follows. Section II describes experimental details, precise steps for experiments, characterization, and instruments involved. All the experimental results are critically scrutinized and discussed in

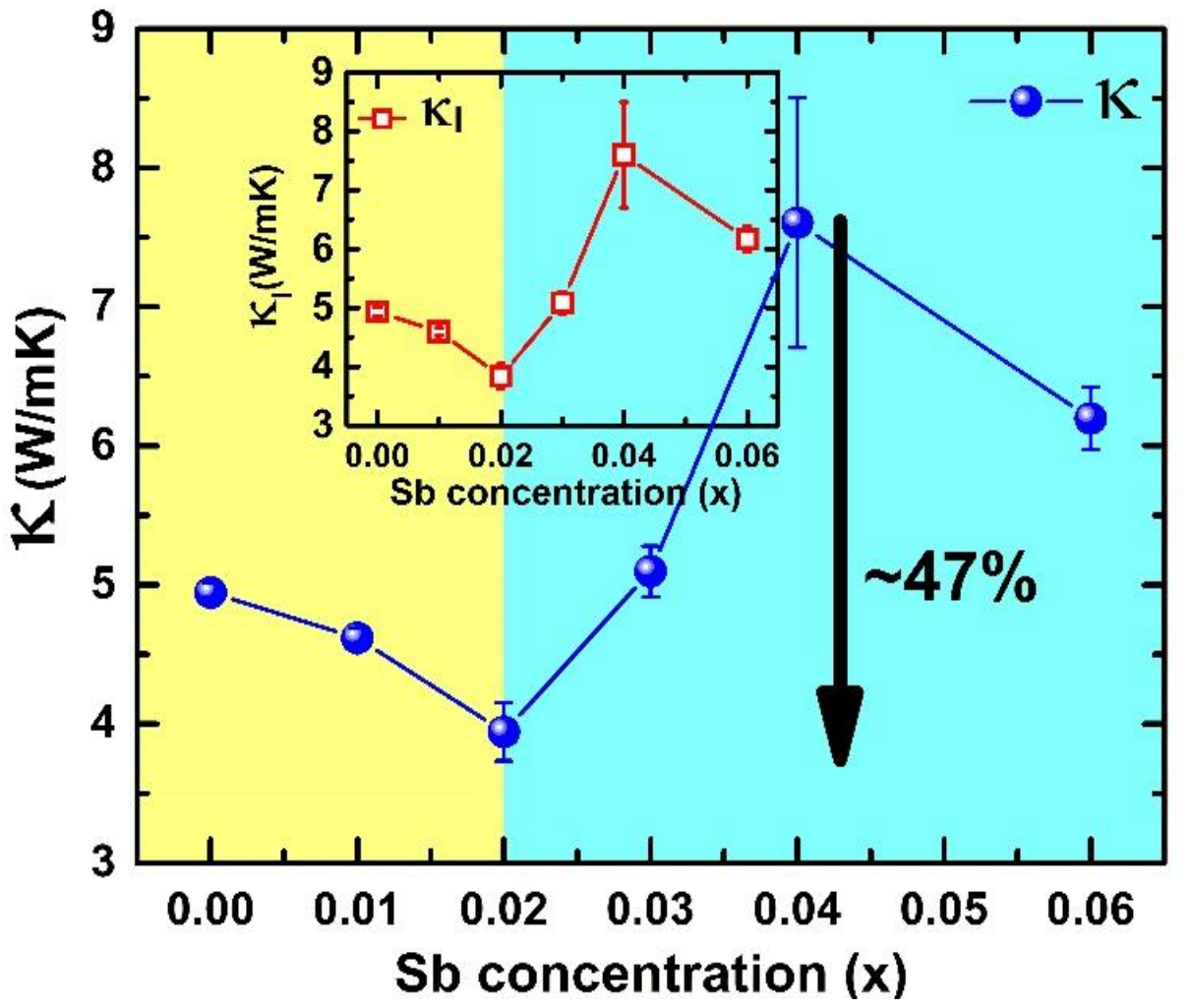

**Fig. 1.** (Color online). Sb concentration dependent thermal conductivity (κ) of synthesized $TiCoSb_{1+x}$ (x=0.0, 0.01, 0.02, 0.03, 0.04, and 0.06) polycrystalline sample. Inset represents the lattice part of thermal conductivity ($\kappa_l$) as a function of Sb concentration. The basic details for the calculation of the electronic and lattice part of thermal conductivity are provided in Supporting Information.

Sec. III. Finally, conclusions and correlations are drawn based on the experimental results in Sec. IV.

## II. Experimental Details

$TiCoSb_{1+x}$ (x=0, 0.01, 0.02, 0.03, 0.04, and 0.06) samples were synthesized by arc melting of the constituent elements. Heat treatment of the ingots, obtained from arc-melting, were performed in a vacuum-sealed quartz tube.[32] The details of synthesis and in-depth structural characterization of the $TiCoSb_{1+x}$ alloys using XRD were reported elsewhere.[32] ZEISS EVO-MA 10 microscope equipped with tungsten filament bulb was used to collect SEM images. The microchemical analysis of synthesized samples was performed by utilising an energy dispersive X-ray (EDX) detector attached to the instrument. In order to carry out SEM, samples were drop-caste onto a glass substrate and allowed to dry at room temperature for three hours. XAFS data of $TiCoSb_{1+x}$ samples were taken at Co K-edge (7708.9 eV) in transmission mode. The details of the working principle of the instrument for XAFS study are explained in Supporting Information (Sec. 1). X-ray photoelectron spectroscopy (XPS) study was employed to investigate the bonding states of the constituent elements in synthesized $TiCoSb_{1+x}$. The Al Kα (1486 eV) X-ray source and an Omicron energy analyser (EA-125) were used to perform XPS measurements. $Ar^+$ ions were used to sputter the surface of the sample for five minutes prior to data collection. The background vacuum in the analyser chamber was maintained at precisely $10^{-10}$ Torr, throughout the measurement. RS measurements (Model: inVia, Make: Renishaw, UK) were carried out at room temperature. The measurements were performed in the range 150-400 $cm^{-1}$, utilizing a 785 nm solid-state laser with a sub-micron focusing diameter (objective of 50x magnification). A 2400gm/cc grating was used for monochromatization with TE-cooled charged coupled device (CCD) as a detector in the back-scattering configuration. Further, temperature and power-dependent RS measurements were carried out to estimate the κ of the synthesized samples. Lattice structure of the synthesized samples at atomic scale were revealed from TEM measurement. TEM measurements were performed using JEM-2100 Plus instrument. Fine powder of the samples was dissolved in ethanol and deposited onto a carbon-coated copper grid to carry out TEM measurements.

## III. Results and Discussion

Phase purity of the synthesized $TiCoSb_{1+x}$ (x=0, 0.01, 0.02, 0.03, 0.04, and 0.06) samples is checked by analysis of the XRD data, and in-depth structural characterizations are performed by Rietveld refinement of XRD data using FullProf software.[32] All the structural parameters obtained after refinement are presented in Table 1 and reported elsewhere.[32] The negligible difference in theoretical and experimental graphs, after the refinement along with other important parameters justify that a good fitting is achieved. κ of synthesized samples is estimated using the structural parameters (Table 1) of the synthesized samples, obtained after Rietveld refinement. The detailed calculation of κ of the synthesized samples is presented elsewhere,[32] and basic calculations to estimate κ are presented in Supporting Information (Sec. 2).

Non-monotonic variation of κ as a function of Sb concentration at room temperature is observed and depicted in

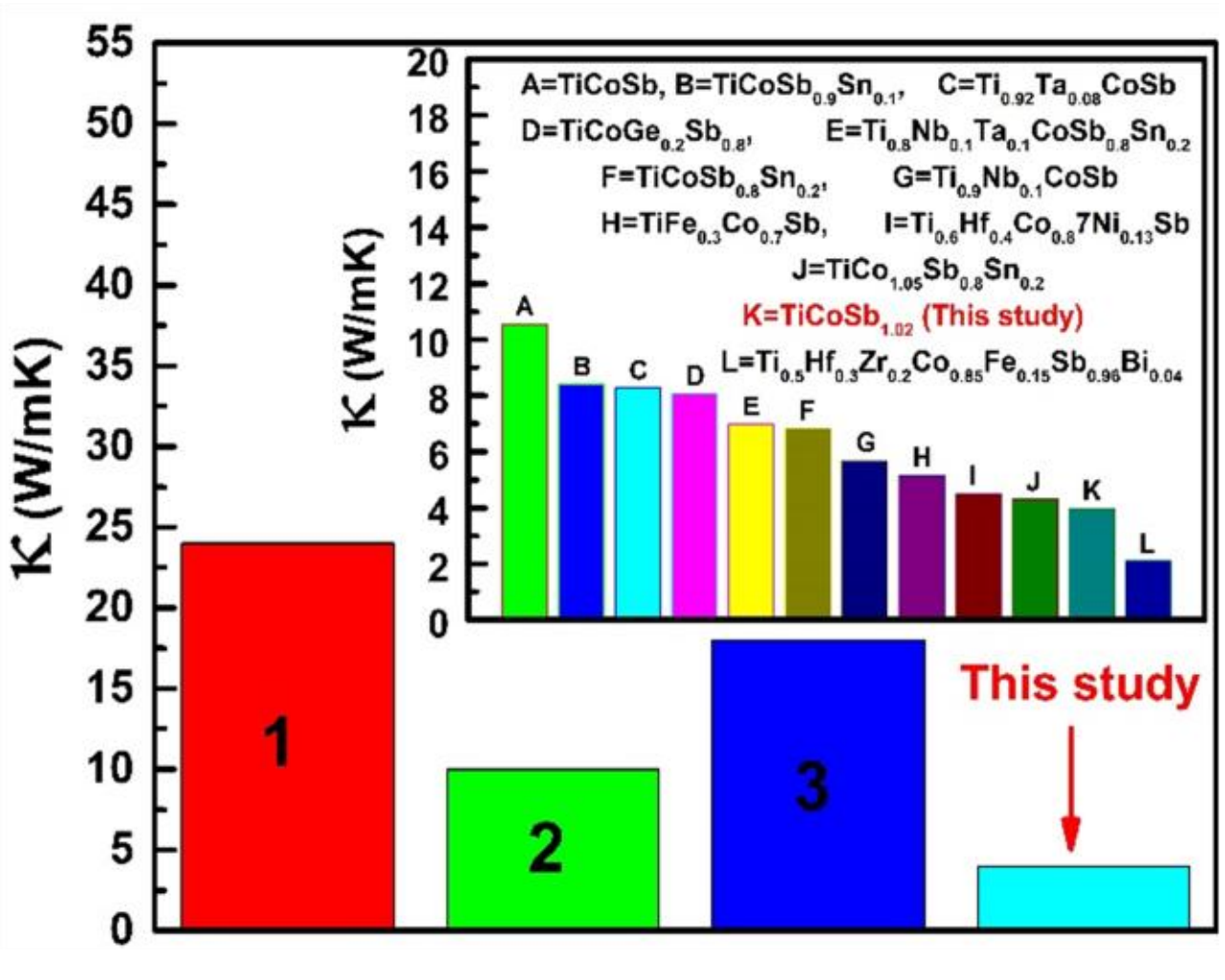

**Fig. 2.** (Color online). A comprehensive comparison of the thermal conductivity of pristine TiCoSb with the reported values in the available literatures (1, 2, 3).[33-35] The inset illustrates the thermal conductivity of numerus doped TiCoSb based HH alloys (A, B, C, D, E, F, G, H, I, J, L).[11, 34-41]

**Table 1.** Structural parameters for synthesized $TiCoSb_{1+x}$ (x=0.0, 0.01, 0.02, 0.03, 0.04, and 0.06) HH polycrystalline alloys, obtained from Rietveld refinement of X-ray diffraction data.[32]

| Sb concentration (x) | Debye-Wallar factor, $B_{iso}$ (Å$^2$) | Debye temperature, $\theta_D$ (K) | Volume (Å$^3$) | Weight percentage (wt%) obtained from XRD | | |
|---|---|---|---|---|---|---|
| | | | | TiCoSb | CoTi | CoSb |
| 0.00 | 0.53±0.01 | 387±2.5 | 203.72±0.01 | 97.40±0.13 | 2.50±0.09 | 0.10±0.01 |
| 0.01 | 0.57±0.01 | 378±2.6 | 203.70±0.01 | 98.49±0.19 | 1.24±0.07 | 0.27±0.02 |
| 0.02 | 0.68±0.03 | 356±6.2 | 203.65±0.02 | 98.76±0.18 | 0.72±0.07 | 0.52±0.02 |
| 0.03 | 0.52±0.02 | 391±6.0 | 203.62±0.01 | 98.63±0.14 | 0.60±0.03 | 0.77±0.03 |
| 0.04 | 0.37±0.04 | 447±18.1 | 203.55±0.03 | 97.94±0.14 | 0.56±0.05 | 1.50±0.05 |
| 0.06 | 0.44±0.02 | 417±6.9 | 203.75±0.04 | 96.40±0.12 | 0.20±0.04 | 3.40±0.07 |

fig. 1. It is crucial to note that there is a wide variation of κ with Sb concentration. The difference between the minimum ($TiCoSb_{1.02}$) and maximum ($TiCoSb_{1.04}$) of κ amid the synthesized samples is ~ 47%. However, $TiCoSb_{1.02}$ contains maximum host phase, i.e., TiCoSb. A comparative chart, using available literature for κ of pristine and doped TiCoSb is presented in fig. 2. *Compared with pristine TiCoSb, κ for our synthesized $TiCoSb_{1.02}$ sample is the minimum to the best of our knowledge. Lowest κ ~3.9 W/mK is achieved for $TiCoSb_{1.02}$ sample*. Further, the minimum value of κ with maximum host phase (TiCoSb) is also comparable with doped TiCoSb [fig. 2 (inset)].

Raman spectroscopy (RS) is an important non-destructive tool to reveal the phonon vibrational modes and is correlated with atomic arrangements in the material. The details of vibration modes and the effect of Sb concentration in TiCoSb HH alloy, obtained from RS data, are discussed in the appropriate sub-section. However, here we only focus on the κ, estimated from temperature and power-dependent RS data. Figure 3 depicts the temperature-dependent κ for the $TiCoSb_{1+x}$ (x=0.0, 0.02, and 0.04) synthesized samples, and corresponding RS data are presented in Supporting Information (fig. S1). Temperature-dependent κ is estimated, employing the Optothermal Raman Technique (ORT),[42, 43] and is compared with the data presented in fig. 1, estimated from XRD data employing Rietveld refinement. In-depth heat conduction processes in a sample due to an incident laser may be explored by RS data.[44] The estimation of κ of a homogeneous bulk sample requires temperature- and power-dependent RS data. This technique is widely used to estimate κ,[42, 43] and it is defined by the following equation, considering radial heat flow [44]

$$\kappa = \frac{\left(\Delta\omega/_{\Delta T}\right)}{4\sqrt{\pi}a\left(\Delta\omega/_{\Delta P}\right)} \tag{1}$$

where $P$, $\left(\Delta\omega/_{\Delta T}\right)$, $\left(\Delta\omega/_{\Delta P}\right)$, and $a$ are incident laser power during the experiment, temperature and power coefficients, and laser spot size (~1.3 μm) respectively. $\kappa$ values obtained from the ORT method for $TiCoSb_{1+x}$ (x=0, 0.02, and 0.04) HH samples are in the same order at room temperature as presented in fig. 3, estimated from structural parameters, obtained by Rietveld refinement of XRD data employing Slack equation.[32] It is crucial to note that minimum κ, estimated from temperature and power-dependent RS data, is also obtained for the $TiCoSb_{1.02}$ sample. The results are elucidated and correlated with other experimental findings in the following sub-sections.

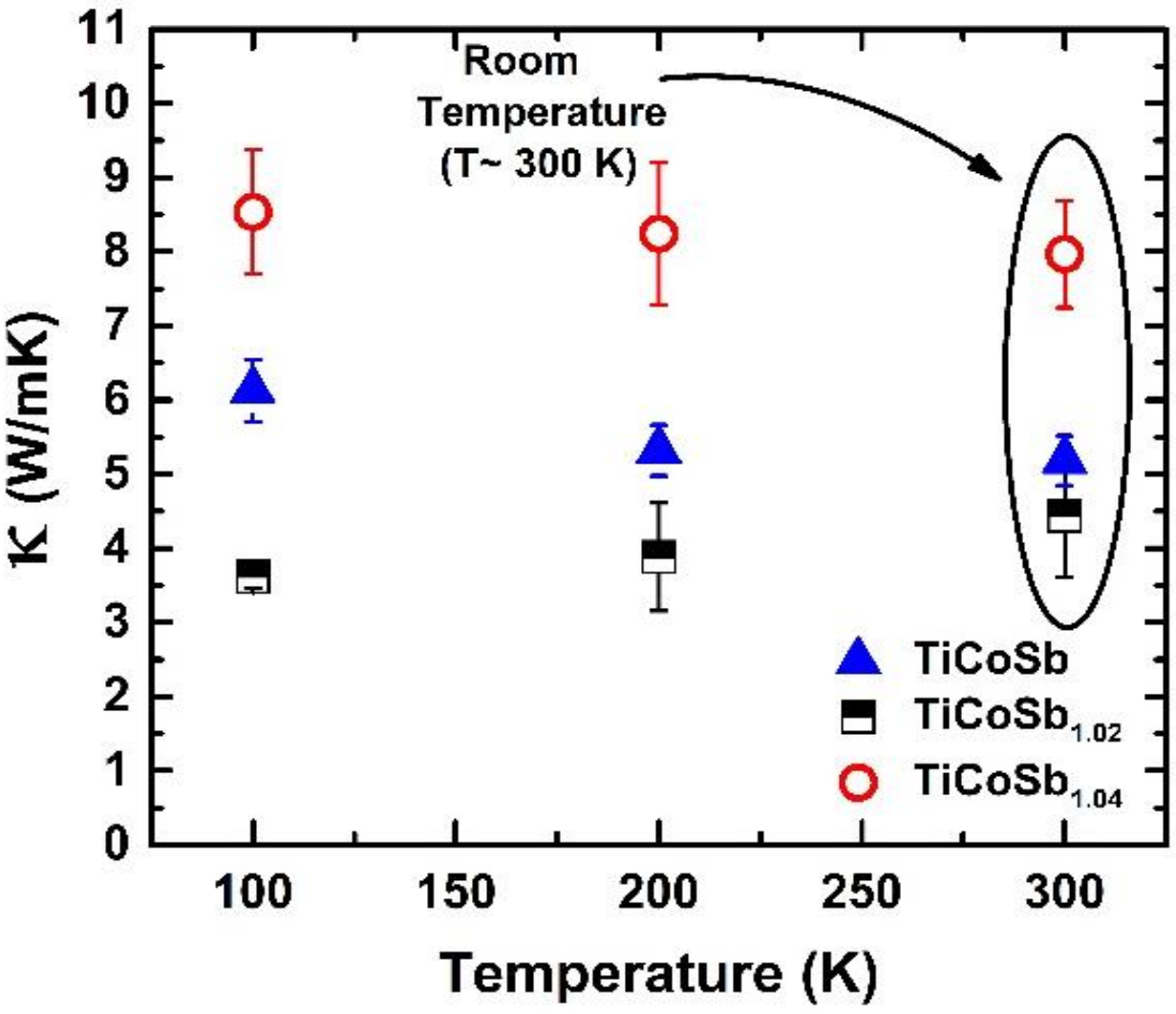

**Fig. 3.** (Color online). Temperature dependent thermal conductivity (κ) of synthesized $TiCoSb_{1+x}$ (x=0, 0.02, and 0.04) half-Heusler alloy, estimated from temperature and power-dependent Raman Spectroscopy data, employing Optothermal Raman Technique (ORT).

### A. Secondary electron microscope and Energy-dispersive X-ray Spectroscopy (EDS)

Microstructural and elemental analysis of all the $TiCoSb_{1+x}$ (x=0, 0.01, 0.02, 0.03, 0.04, 0.06) HH alloys are carried out using SEM and EDS. Figure 4 depicts the SEM micrograph and results of EDS analysis are presented in Supporting Information (fig. S2). Figure 4 reveals that the shining spot within the TiCoSb matrix (dark area) may be related to the presence of a minute amount of embedded phases within the TiCoSb matrix (Table 1). Rietveld refinement of XRD data reveals the presence of metallic phases in the TiCoSb matrix of the synthesized samples. The effect of metallic phases in the host matrix is also observed in transport properties of the synthesized materials.[32] Metallic phases typically exhibit a shiny appearance. Shiny Co metallic phases are also observed in the SEM picture of $Ba_{0.3}In_{0.3}Co_4Sb_{12}$ TE material.[45] The detailed analysis of EDS data is given in Supporting Information (Sec. 4). EDS data indicates a slight deviation in atomic percentages from the designed values (Supporting Information), suggest the formation of embedded phases within host matrix. Attempt is taken to estimate the stoichiometry of the synthesized $TiCoSb_{1+x}$ samples along with embedded phases (Table 1) using atomic percentages, obtained from EDS results (Table 2). It is significant to mention that the weight percentage (wt%) of embedded phases i.e., CoTi and CoSb, derived from the EDS data are nicely matched with the Rietveld refinement data (Table 1). The presence of these phases are also identified through a critical analysis of the XRD patterns, utilizing mixed-phase Rietveld refinement.[32] The SEM and EDS results confirm the presence of embedded metallic phases and Co vacancies in the TiCoSb matrix.

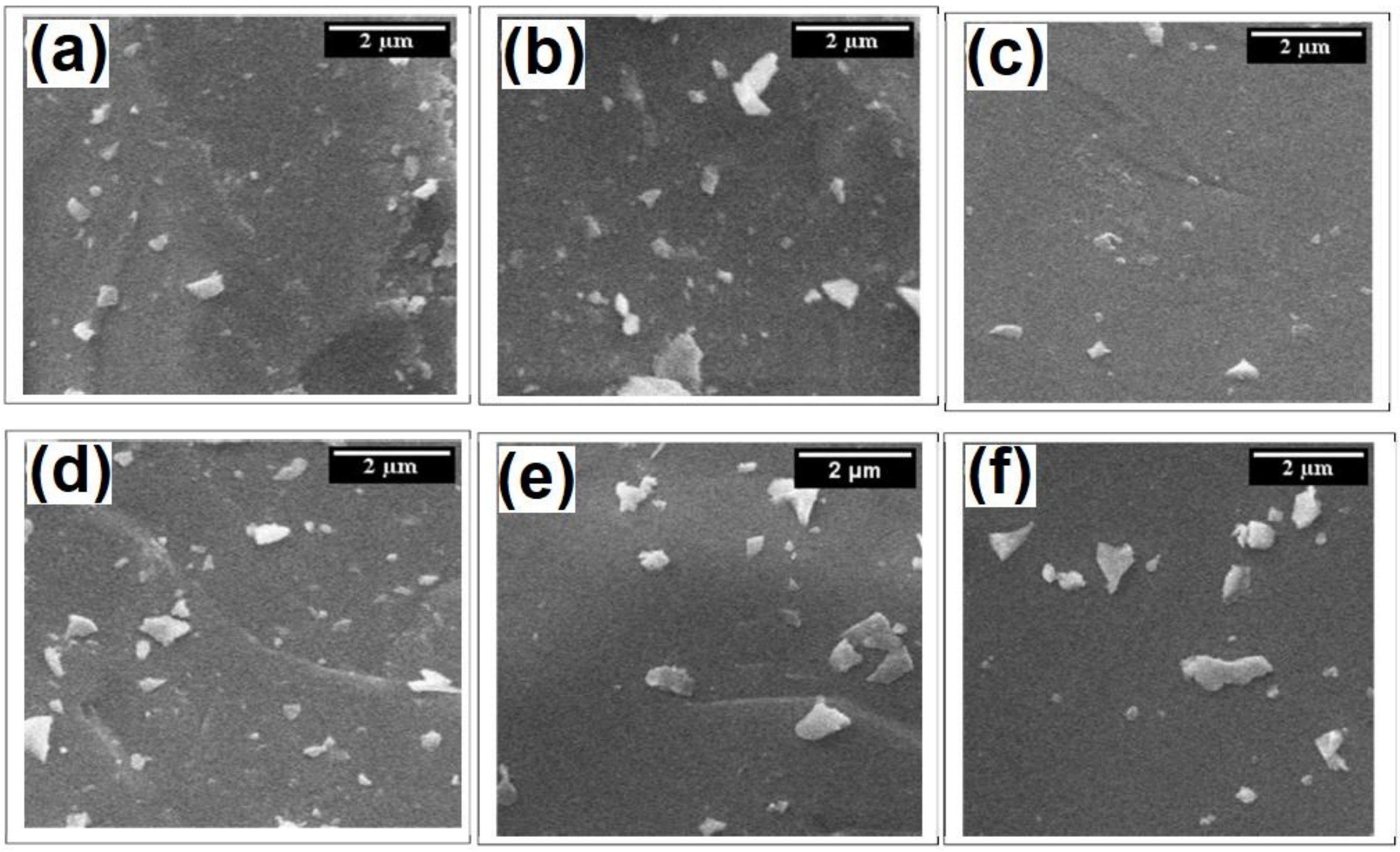

**Fig. 4.** (Color online). Scanning electron microscopic images of the synthesized $TiCoSb_{1+x}$ samples for a) x=0.0, b) x=0.01, c) x=0.02, d) x=0.03, e) x=0.04, and f) x=0.06. Distinguishable variation in TiCoSb matrix is evidence of the embedded phases in the synthesized alloy.

**Table 2.** Weight percentage (wt%) of embedded phases in synthesized samples, derived from the EDS result of the synthesized $TiCoSb_{1+x}$ (x=0.0, 0.01, 0.02, 0.04, and 0.06) HH polycrystalline alloys. The errors in the data are within ±0.09.

| Sb concentration (x) | Wt% derived from the EDS result | | | Co vacancy, $V_{Co}$ (%) |
|---|---|---|---|---|
| | TiCoSb | CoTi | CoSb | |
| 0.00 | 97.00 | 3.00 | 0.00 | 0.00 |
| 0.01 | 98.56 | 1.46 | 0.00 | 0.50 |
| 0.02 | 99.10 | 0.90 | 0.00 | 0.70 |
| 0.04 | 98.22 | 0.00 | 1.78 | 2.00 |
| 0.06 | 96.50 | 0.00 | 3.50 | --- |

## B. Transmission electron microscopy (TEM)

High resolution TEM (HRTEM) measurements of $TiCoSb_{1+x}$ (x = 0, 0.02, and 0.04) HH alloys are conducted for in-depth structural characterization. The TEM images and corresponding selected area electron diffraction (SAED) patterns are presented in fig. 5. The average particle size of synthesized samples are the order of a few hundred nanometers (figs. 5(a-c)). Maximum particle size is obtained for $TiCoSb_{1.02}$ samples. HRTEM images of crystallographic planes are presented in figs. 5(d-f). The interplanar spacing of the crystallographic planes of HH alloys are estimated from the TEM images and presented in insets of the figs. 5(d-f). It is crucial to note that the lattice spacing slightly decreases for the $TiCoSb_{1.02}$, and $TiCoSb_{1.04}$ samples, indicates a contraction in lattice volume with increasing Sb concentration.[32] The SAED patterns of the $TiCoSb_{1+x}$ (x=0, 0.02, and 0.04) samples as depicted in figs. 5(g-i), are utilized to estimate the crystallographic planes in the synthesized $TiCoSb_{1+x}$ HH alloys. The perfect ring patterns confirm the good crystalline nature of the synthesized samples. The estimated crystallographic planes are also observed in the XRD pattern of the $TiCoSb_{1+x}$ samples (x=0, 0.01, 0.02, 0.03, 0.04, and 0.06).[32] Further, the absence of additional diffraction rings or spots indicates the phase purity of the synthesized samples.[32] Interplanar spacing or d-spacing is also estimated from the SAED pattern and presented in Supporting Information (Table S1). It is important to note that d-spacing corresponding to the (h k l), obtained from HRTEM images (figs. 5(d-f)) and SAED pattern (figs. 5(g-i)), decreases with an increase in Sb concentration (x), confirms the contraction in lattice volume.

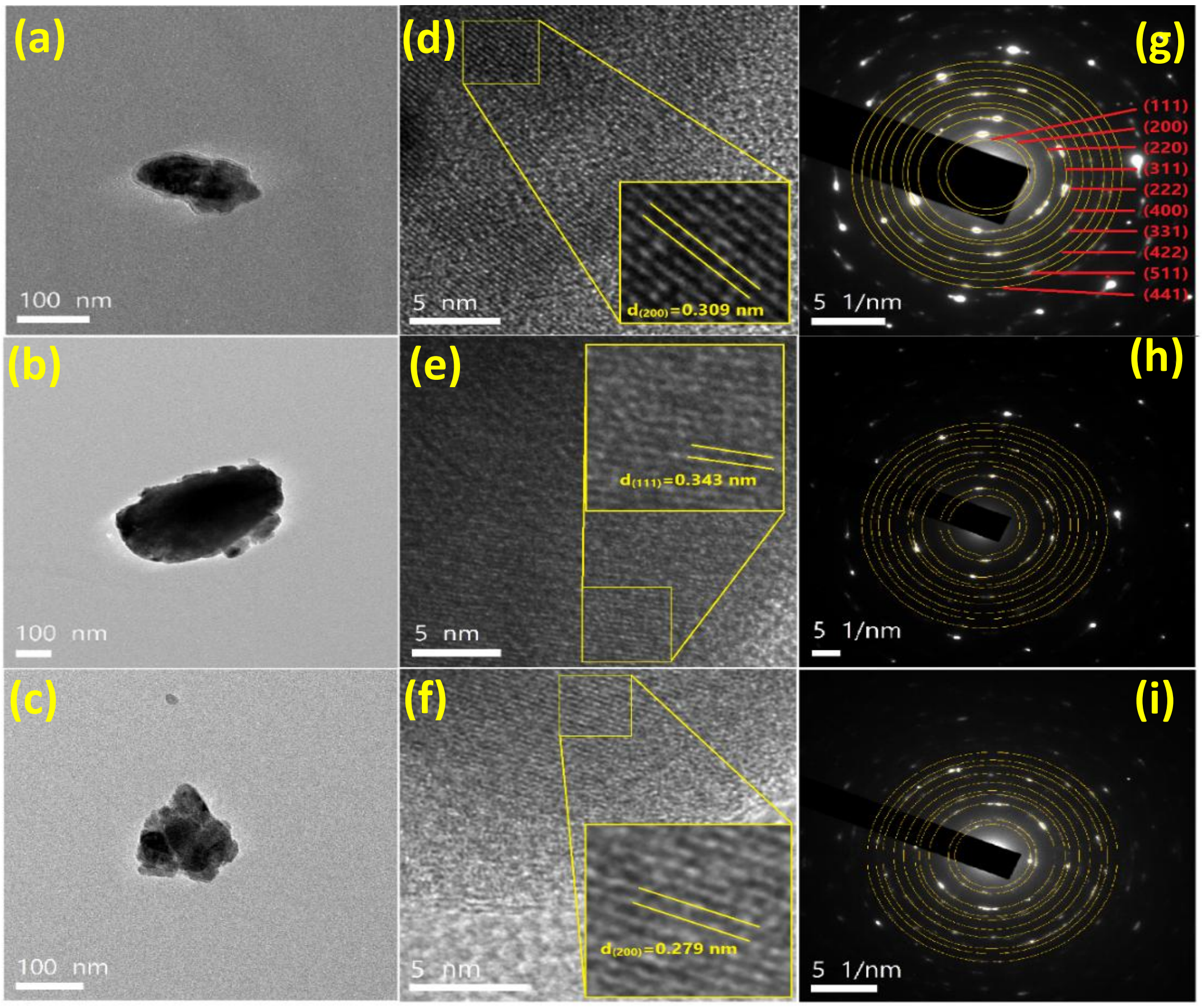

**Fig. 5.** (Color online). Transmission electron microscopy (TEM), high-resolution TEM (HRTEM), and selected area electron diffraction (SAED) pattern of the synthesized $TiCoSb_{1+x}$ alloy. (a-c) TEM, (d-f) HRTEM, and (g-i) SAED patterns for x = 0.0, 0.02, and 0.04, respectively. The insets of HRTEM images display magnified views of the lattice planes. The rings in the SAED patterns correspond to the (h k l) planes of the TiCoSb half-Heusler (HH) phase.

### C. X-ray absorption fine structure spectroscopy: Investigation of short range order

X-ray absorption fine structure (XAFS) spectroscopy using a tunable photon source, viz. the synchrotron radiation, is used to obtain the information regarding atomic coordination around a particular atom (viz. crystalline order or disorder, atomic defect) of a system.[46] The XAFS data has two parts: (i) X-ray absorption near edge structure (XANES) spectroscopy, which reflects electronic features (e.g., valance state) and is strongly related to coordination atoms, and (ii) EXAFS spectroscopy reflects the local arrangement of neighbouring atoms for a wide range.[25] Room temperature XAFS data of synthesized samples, obtained in the range 7600 eV to 8500 eV, are given in Supporting Information (fig. S3). We have elaborately discussed the results sequentially.

#### i. X-ray absorption near edge structure (XANES)

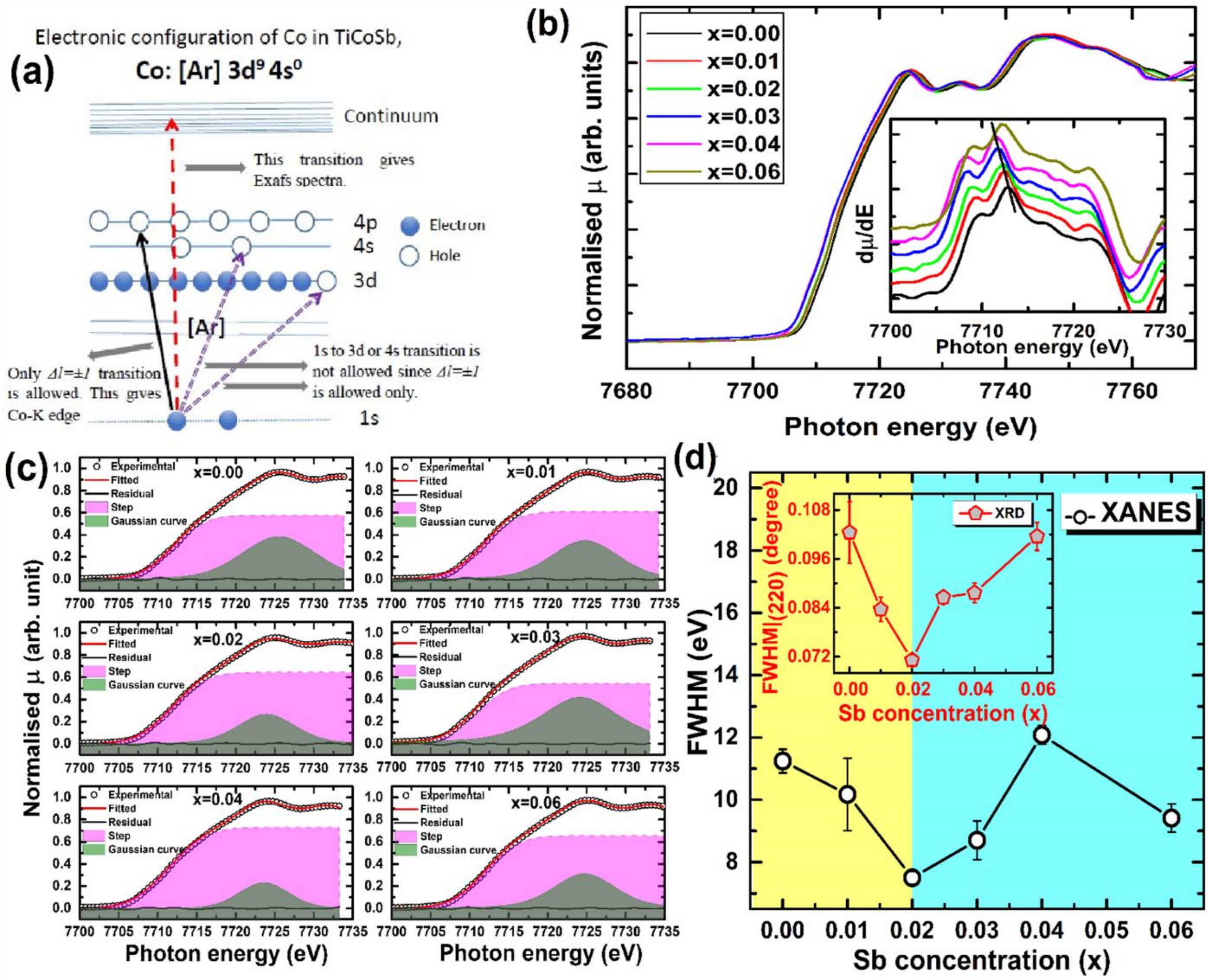

**Fig. 6.** (Color online). (a) Allowed transition for Co K-edge spectrum. (b) Normalized XANES spectra of synthesized $TiCoSb_{1+x}$ polycrystalline samples at Co-K edge. Inset shows the first derivative of the edge jump. (c) Fitted XANES spectra for each $TiCoSb_{1+x}$ samples at Co-K edge. (d) Full Width at Half Maxima (FWHM) as a function of Sb concentration, obtained from XANES spectra. Inset represents the variation of FWHM with Sb concentrations, obtained from X-ray diffraction of $TiCoSb_{1+x}$ ($0.0 \leq x \leq 0.06$) alloy.

XANES region in Co K-edge spectrum is observed due to the excitation of the 1s electron to the unoccupied 4p level (fig. 6(a)). It is important to note that, amid all possible transitions, s→d (1s→3d) and s→s (1s→4s) transitions are forbidden owing to the quantum selection rules, i.e., only $\Delta l = \pm 1$ transition is allowed [fig. 6(a)].[47] Hence, the 1s→4p transition becomes the only pathway to excite 1s electrons to the higher levels [fig. 6(a)] for XAFS at Co K-edge of TiCoSb. Figure 6(b) depicts the XANES spectra of $TiCoSb_{1+x}$ ($0.0 \leq x \leq 0.06$) at Co K-edges. The first derivative of XANES spectra is presented in the inset of fig. 6(b). A significant shift of the Co K-edge towards lower photon energy is observed [inset fig. 6(b)] with the addition of Sb concentration, manifests the change in the Co oxidation state.[48, 49] A shift ~2 eV is obtained owing to the change in the average oxidation state of Co in $TiCoSb_{1+x}$ samples for $0 \leq x \leq 0.04$. The shift towards the lower photon energy is related to the screening effect due to a decrease in the average valence state of Co ions, i.e., an increase in the overall negative charge of the atom.[48, 49] The change in oxidation states in the synthesized $TiCoSb_{1+x}$ HH alloy may be related to Co vacancy. Co vacancy gradually increases with increasing Sb concentration in $TiCoSb_{1+x}$ alloy. The result exactly reflects in Rietveld refinement (Table 1) and EDS (Table 2) data. In order to investigate Co K-edge spectra quantitatively XANES region is fitted with one arctangent step function and a pseudo-voigt peak shape function [fig. 6(c)], accommodating the electronic transition from 1s to continuum states and the transition to the unoccupied states, respectively. The fig. 6(d) represents the full width at half maxima (FWHM) for the white line of Co K-edge XANES spectra of $TiCoSb_{1+x}$ $(0 \leq x \leq 0.06)$ synthesized samples. The FWHM of the white line decreases up to x= 0.02, indicates the decrease in local disorder around the Co atom. However, FWHM increases for $0.02 < x \leq 0.06$ samples is the manifestation of an increase in disorder compared to x=0.02.[50] X-ray diffraction (XRD) data also reveal that disorder is minimum and the best crystal quality is obtained for $TiCoSb_{1.02}$ sample.[32] Hence, the structural model considered for the analysis of XANES results is in good agreement with the long range order, investigated by the XRD measurement.[32]

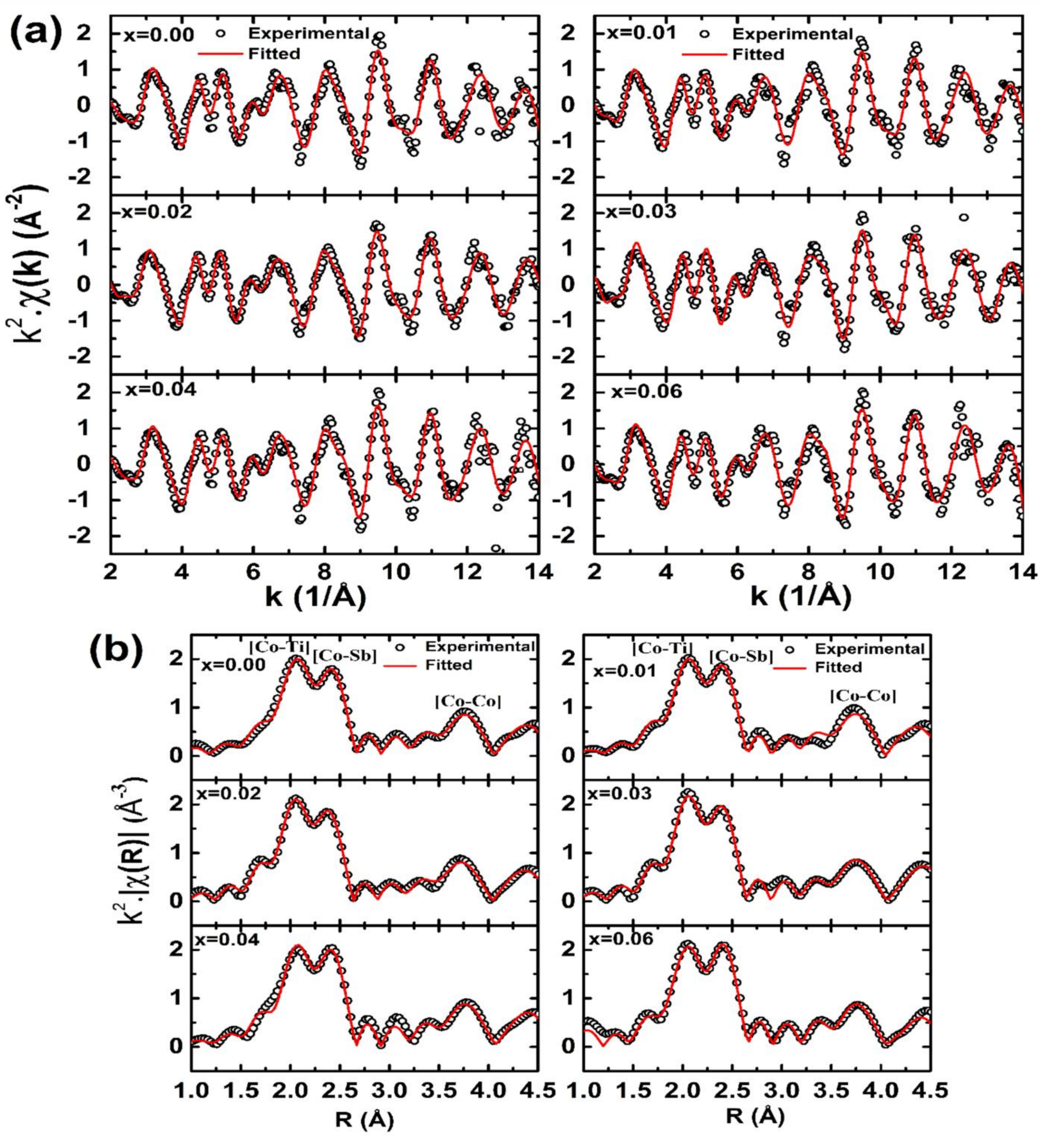

**Fig. 7.** (Color online). (a) $k^2$ -weighted derivatives, and (b) radial distributions of Co K-edge X-ray absorption spectra, surrounding Co atom of the synthesized $TiCoSb_{1+x}$ (x=0.0, 0.01, 0.02, 0.03, 0.04, and 0.06) HH samples, (open circles and red- lines are the experimental and fitted curves, respectively).

**Table. 3.** Degeneracy, scattering length, rank, and type of scattering are presented for all probable scattering paths at Co k-edge of the synthesized $TiCoSb_{1+x}$(x=0.0, 0.01, 0.02, 0.03, 0.04, and 0.06) HH samples, used during FEFF modeling. Highlighted paths are included in the analysis during fitting.

| Scattering Path at Co-k edge | Degeneracy N | Scattering length, $R_{eff}$ (Å) | Rank | Type of Scattering |
|---|---|---|---|---|
| **@ Sb8.1 @** | **4.00** | **2.552** | **100** | **Single** |
| **@ Ti3.1 @** | **4.00** | **2.552** | **80.09** | **Single** |
| @ Sb8.1 Ti3.1 @ | 24.00 | 4.025 | 10.36 | Other double |
| **@ Co5.1 @** | **12.00** | **4.167** | **64.65** | **Single** |
| @ Sb8.1 Sb10.1 @ | 12.00 | 4.636 | 6.63 | Other double |
| @ Sb8.1 Co5.1 @ | 24.00 | 4.636 | 20.31 | Other double |
| @ Ti3.1 Ti0.1 @ | 12.00 | 4.636 | 7.55 | Other double |
| **@ Ti3.1 Co5.1 @** | **24.00** | **4.636** | **20.69** | **Other double** |
| **@ Sb8.2 @** | **12.00** | **4.887** | **63.27** | **Single** |
| **@ Ti3.2 @** | **12.00** | **4.887** | **51.26** | **Single** |

### ii. Extended X-ray absorption fine structure (EXAFS)

The EXAFS data of the synthesized samples are depicted in Supporting Information (fig. S3). The oscillation part of the measured absorption coefficient [μ(E)] is converted to fine structure-function ($\chi(k)$) in k-space.[51] Detail procedure of data conversion is provided in Supporting Information (Sec. 6). $\chi(k)$ is weighted by $k^2$ to amplify oscillations at high k, and corresponding fittings for the synthesized samples are presented in fig. 7(a). The $k^2\chi(k)$ function is Fourier transformed to real space (R-space) i.e., $k^2\chi(k)$ employing the ARTEMIS software package (Supporting Information).[52] In order to fit the experimental data, scattering paths are estimated by FEFF modeling with integration of IFEFFIT employing ARTEMIS.[52] The scattering paths, included in fittings, are highlighted in Table 3, and detail explanation is provided in Supporting Information (Sec. 6). The theoretical fittings of the R-space data for the synthesized samples are depicted in fig. 7(b) and a schematic for visualization of the scattering effect is shown in fig. 8. The Fourier transforms show well-defined peaks around 2.5 and 4.2 Å [fig. 7(b)]. These correspond to the single scattering in the nearest neighbour shell, i.e., coordination shell, and second nearest neighbour shell, respectively. The absorber Co atom is coordinated tetrahedrally by four Sb and Ti atoms (Reff = 2.5 Å) and dodecahedral by 12 Co (Reff = 4.167 Å) atoms. Crucial to mention that, a splitting of the first peak is observed in the Fourier transformed (R = 2-2.5 Å) EXAFS data [fig. 7(b)], due to different scattering phases of Ti and Sb atoms. Small intense peaks in between three main peaks, i.e., $2.5\ Å \leq R \leq 3.5\ Å$ region, indicate the contribution of multiple scattering [fig. 7(b)]. The reliability of the fits is asserted by the small values of the $\sigma^2$ and R-factors (Table 4). Nevertheless, the comparison of the fitted and experimental data in fig. 7(a) and fig. 7(b) also confirms the goodness of the fits. The EXAFS data analysis opens the path to investigate the effect of Sb concentration on the local geometry of $TiCoSb_{1+x}$ HH alloy. The effect of Sb concentration on the bond length is also estimated and represented in Table 4. A significant change in Co-Ti bond length ($R_{Co\text{-}Ti}$) of the 1st shell and a minute change in Co-Sb bond length ($R_{Co\text{-}Sb}$) of the 2nd shell are observed. $R_{Co\text{-}Ti}$ and $R_{Co\text{-}Sb}$ of the synthesized $TiCoSb_{1+x}$ samples are changed from 2.541 Å to 2.489 Å and 2.543 Å to 2.532 Å, respectively for $0 < x \leq 0.06$. The EXAFS results suggest that the volumes of 1st and 2nd shells are reduced with increasing Sb concentration in $TiCoSb_{1+x}$ owing to Co vacancy in the synthesized samples. Further, the Debye-Waller factors ($\sigma^2$), obtained from fittings, are lower for 2nd shell [$\sigma^2_{Co-Sb} \sim 0.3 \times 10^{-3} Å^2$] than the 1st shell [$\sigma^2_{Co-Ti} \sim 1 \times 10^{-2} Å^2$].

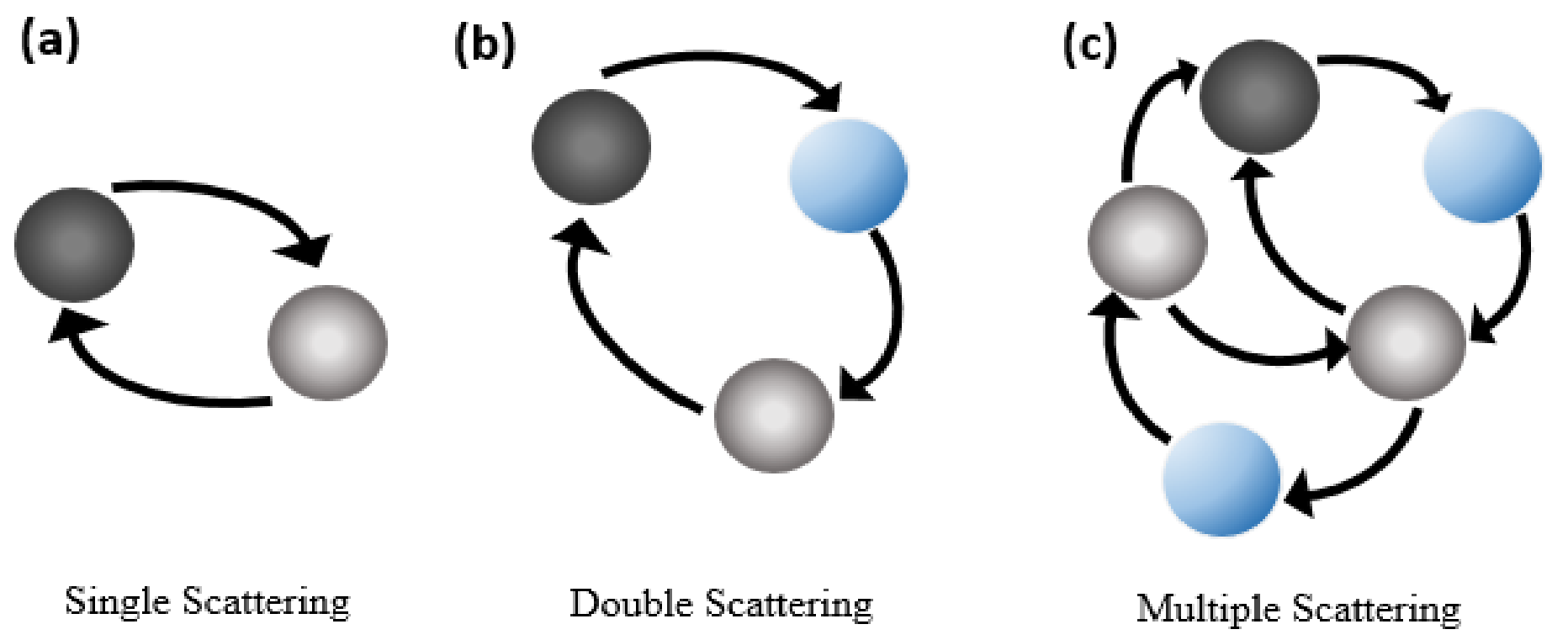

**Fig. 8.** (Color online). Schematic illustrations for the sequential construction of different types of scattering effect, where (a) single scattering of an electron between two atoms, (b) double scattering of an electron involving three atoms, and (c) multiple scattering of an electron among multiple atoms. Atoms are represented by spheres of different colors.

**Table 4.** Theoretical fitting parameters of experimental EXAFS data for the synthesized $TiCoSb_{1+x}$ (x=0.0, 0.01, 0.02, 0.03, 0.04, and 0.06) HH alloys. EXAFS data are taken at the Co K-edge. $R_{Co\text{-}Sb}$, $R_{Co\text{-}Ti}$, $\sigma^2_{Co-Sb}, \sigma^2_{Co-Ti}$, and Happiness indicate Co-Sb bond length, Co-Ti bond length, Debye-Waller factors for 1st and 2nd shell, and Happiness of the fits, respectively.

| Sb Concentration (x) | $R_{Co\text{-}Sb}$ (Å) | $R_{Co\text{-}Ti}$ (Å) | $\sigma^2_{Co-Sb}$ | $\sigma^2_{Co-Ti}$ | Happiness |
|---|---|---|---|---|---|
| 0.00 | 2.543±0.004 | 2.541±0.008 | 0.0034±5.3989E-4 | 0.0104±0.00084 | 100.0 |
| 0.01 | 2.539±0.003 | 2.530±0.006 | 0.0029±5.2486E-4 | 0.0110±0.00105 | 95.77 |
| 0.02 | 2.534±0.003 | 2.506±0.005 | 0.0030±4.7079E-4 | 0.0086±0.00261 | 97.00 |
| 0.03 | 2.532±0.004 | 2.489±0.011 | 0.0025±5.3469E-4 | 0.0106±0.00107 | 98.45 |
| 0.04 | 2.542±0.003 | 2.486±0.006 | 0.0027±5.0786E-4 | 0.0110±0.00104 | 93.14 |
| 0.06 | 2.538±0.005 | 2.537±0.009 | 0.0026±6.1525E-4 | 0.0110±0.00115 | 89.44 |

## D. X-ray Photoelectron Spectroscopy

X-ray photoelectron spectroscopy (XPS) measurements on $TiCoSb_{1+x}$ (x=0, 0.02, and 0.04) HH polycrystalline alloys are performed to reveal the variation in bonding states. Figure 9(a) depicts the survey scan spectrum of the TiCoSb (x=0.0) alloy (spectra for x=0.02 and 0.04 are given in Supporting Information (fig. S4)). Survey scan spectrum possesses photoemission peaks of Ti, Co, and Sb.[53] However, the presence of peaks for carbon (C) and oxygen (O) may be related to absorbed impurities, originated due to atmospheric exposure of the synthesized samples in the course of transfer into the measurement chamber.[54, 55] In order to explore the details of environment and variation in binding energy, photoemission spectra for Ti, Co, and Sb of synthesized $TiCoSb_{1+x}$ (x=0, 0.02, 0.04) samples are also recorded. The typical core-level spectra of Ti, and Sb for the synthesized TiCoSb alloy are depicted in figs. 9(b) and 9(c). Ti and Sb core-level spectra of $TiCoSb_{1.02}$ and $TiCoSb_{1.04}$ are given in Supporting Information (fig. S4). All the Ti, Co, and Sb core spectra are deconvoluted employing XPSPEAK41 software. Ti 2p photoemission spectra of the synthesized samples consist of three distinct peaks at 455 eV, 459 eV, and 464 eV. The peaks at 455 eV in the samples arise due to the presence of metallic bonding of the Ti element in the synthesized samples, whereas other two peaks correspond to $2P_{3/2}$ (~459 eV) and $2P_{1/2}$ (~464 eV) states, respectively.[54] Difference in binding energy position of Ti $2p_{3/2}$ peak (~459 eV) of synthesized samples is ~3.5 eV, compared to the photoemission peak of pure Ti metal (~455.5 eV).[53] Sb 3d core-level spectrum is deconvoluted and depicted in fig. 9(c), consist of three peaks at ~526.5 eV, ~530 eV, and 536.5 eV, respectively. The peaks of the Sb core-level spectrum, 526.5 eV and 536.5 eV, may be assigned to Sb $3d_{3/2}$, and Sb $3d_{5/2}$ states respectively.[54] However, peak at ~530 eV in fig. 9(c) arises likely due to the surface oxidation during measurement of the samples.[54] The binding energy of Sb $3d_{3/2}$ in the synthesized alloys is measured at 536.5 eV, approximately 1.2 eV lower than the position of the $3d_{3/2}$ peak for elemental Sb ~537.8 eV.[54] However, a prominent change is observed in the Co core spectra of the samples (x=0.0, 0.02, and 0.04). The Co 2p core-level spectra of the synthesized samples (x=0.0, 0.02, and 0.04) are presented in fig. 9(d-f). Two peaks are obtained at binding energies of ~778eV and ~792eV, correlated to Co $2p_{1/2}$ and Co $2p_{3/2}$.[55] The prominent peaks (~778eV and ~792eV) for synthesized alloys are at slightly higher values of binding energy concerning the elemental Co atom.[56] However, deconvolution of the Co $2p_{3/2}$ state reveals the presence of two component peaks at binding energies of ~778 eV and 782 eV, designated as A and B in figs. 9(d-f). The presence of A and B indicates mixed oxidation states of Co in the synthesized $TiCoSb_{1+x}$ alloys. The peak A at a lower binding energy state and peak B at a higher binding energy state correspond to lower and higher oxidation states of Co in the synthesized alloys.[55] In order to estimate the area under the curve of A and B, the Co $2p_{3/2}$ peaks are fitted using a Lorentzian peak function following Shirley-type background correction. The ratios of area under the curves of B to A ($B_{area}/A_{area}$) for the synthesized TiCoSb, $TiCoSb_{1.02}$, and $TiCoSb_{1.04}$ alloys are depicted in fig. 9(g). Importantly, the $B_{area}/A_{area}$ increases with Sb concentration for the synthesized alloys.

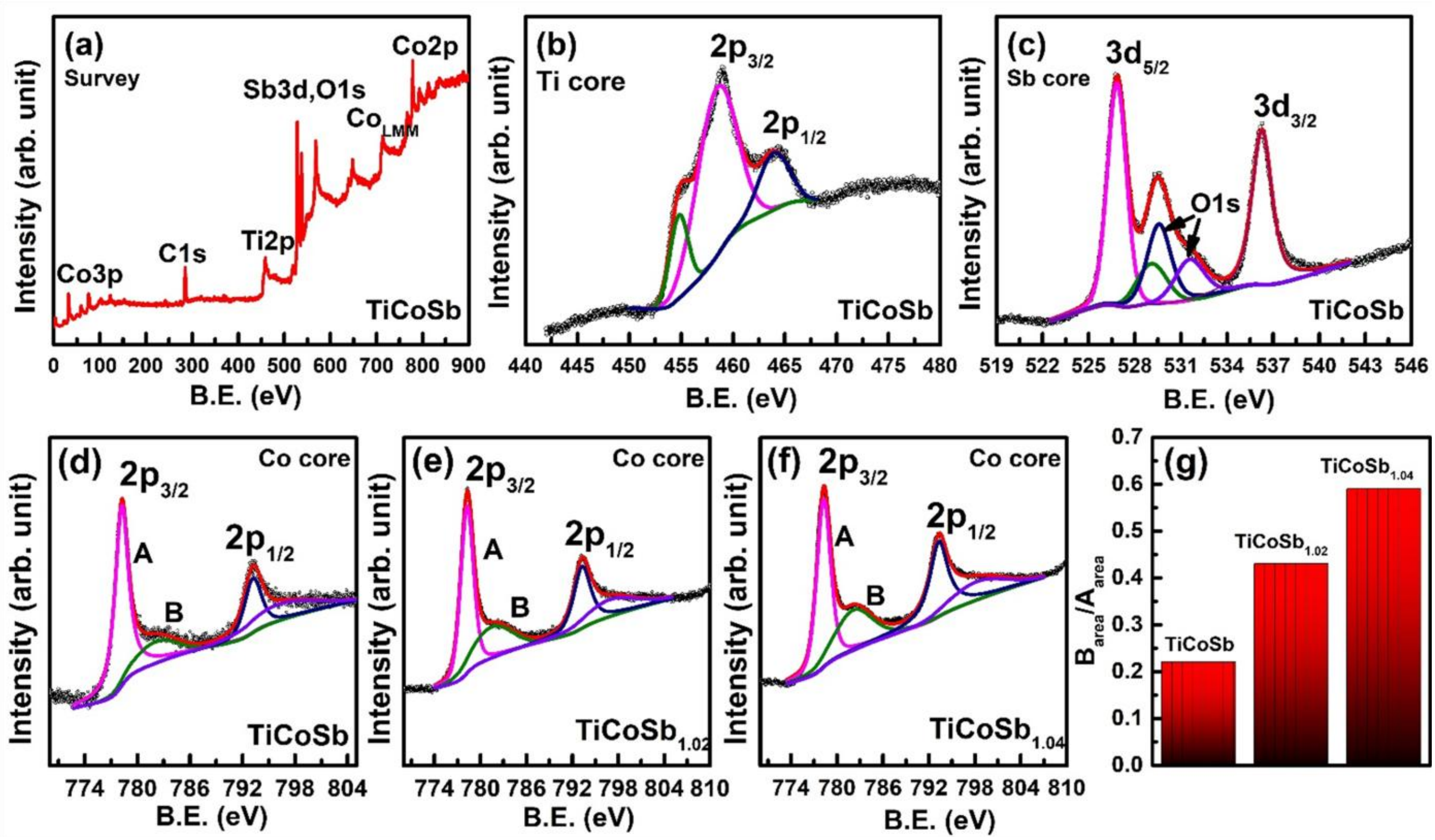

**Fig. 9.** (Color online). X-ray photoelectron spectroscopy (XPS) analysis of $TiCoSb_{1+x}$ HH polycrystalline materials. (a) Survey spectrum of synthesized TiCoSb (x=0.0) alloy. High-resolution XPS spectrum (black dot) accompanying deconvoluted peaks are indicated using color lines (magenta, olive, violet, and navy blue), and Shirley background of the (b) Ti 2p, (c) Sb 3d, and (d) Co 2p core level for the synthesized TiCoSb HH alloy are depicted. The core level spectra of Co 2p are shown in (e) $TiCoSb_{1.02}$, and (f) $TiCoSb_{1.04}$. The peaks indicated by A and B in (d), (e), and (f) related to lower and higher oxidation states of the Co atom, respectively. (g) Quantitative analysis of the ratio of area under the curve of B to A for $TiCoSb_{1+x}$ (x=0.0, 0.02, 0.04) alloys illustrates the systematic variations in the electronic properties with Sb concentration.

It is crucial to note that there is no significant change in the spectra of Ti and Sb core, indicate the environment and bonding states remain similar for the synthesized samples. The presence of metallic bonding in Ti core spectrum may arise due to the CoTi metallic phases in synthesized $TiCoSb_{1+x}$ alloys. The presence of CoTi embedded phases is also observed in Rietveld refinement analysis of the synthesized samples.[32] The shifts in binding energies of Ti $2p_{3/2}$ and Sb $3d_{3/2}$ concerning the position of corresponding elemental atoms, indicate the formation of HH alloys. A minute observation reveals that the shift in binding energy of Ti, Co, and Sb core spectra of synthesized alloys, compared to the corresponding elementals, is not in the same direction. A comparative analysis of chemical shifts indicates opposite trends for Ti and Co relative to Sb, evince charge transfer towards the Sb atom from the Ti and Co atoms. The charge transfer mechanism, observed in XPS spectra of synthesized alloys, well aligns with the characteristics of XYZ type HH alloys. X and Y elements are more electropositive than Z elements in this type of alloys, results X and Y exhibit cationic character, and Co behaves as anion.[57]

### E. Raman spectroscopy measurements

A comprehensive analysis of the vibrational spectroscopy of the synthesized $TiCoSb_{1+x}$ HH alloys is carried out using RS measurement at room temperature. RS data of $TiCoSb_{1+x}$ samples is depicted in fig. 10(a). No evidence of embedded phases is observed within the measured frequency range. The experimental spectra of synthesized samples exhibit three prominent peaks located at 230 $cm^{-1}$, 264 $cm^{-1}$, and 318 $cm^{-1}$ [fig. 10(a)]. There are no reported experimental RS data of TiCoSb HH alloy to the best of our knowledge. However, Popovic et al. and Mestres et al. have designated that the first two peaks are associated with TO-modes and the third peak as a LO-mode for similar types of HH alloy.[58] Computational analysis by Hermet et al. unequivocally confirms the identification of TO2 and LO2 modes at around 264 $cm^{-1}$ and 318 $cm^{-1}$ for NiTiSn based HH alloy.[59] The intrinsic scattering efficiency of LO1 and TO1 modes is comparable.[59] Hence, LO1 and TO1 may overlap with each other in RS data.[59] Hence, the first peak at 228 $cm^{-1}$ for RS of $TiCoSb_{1+x}$

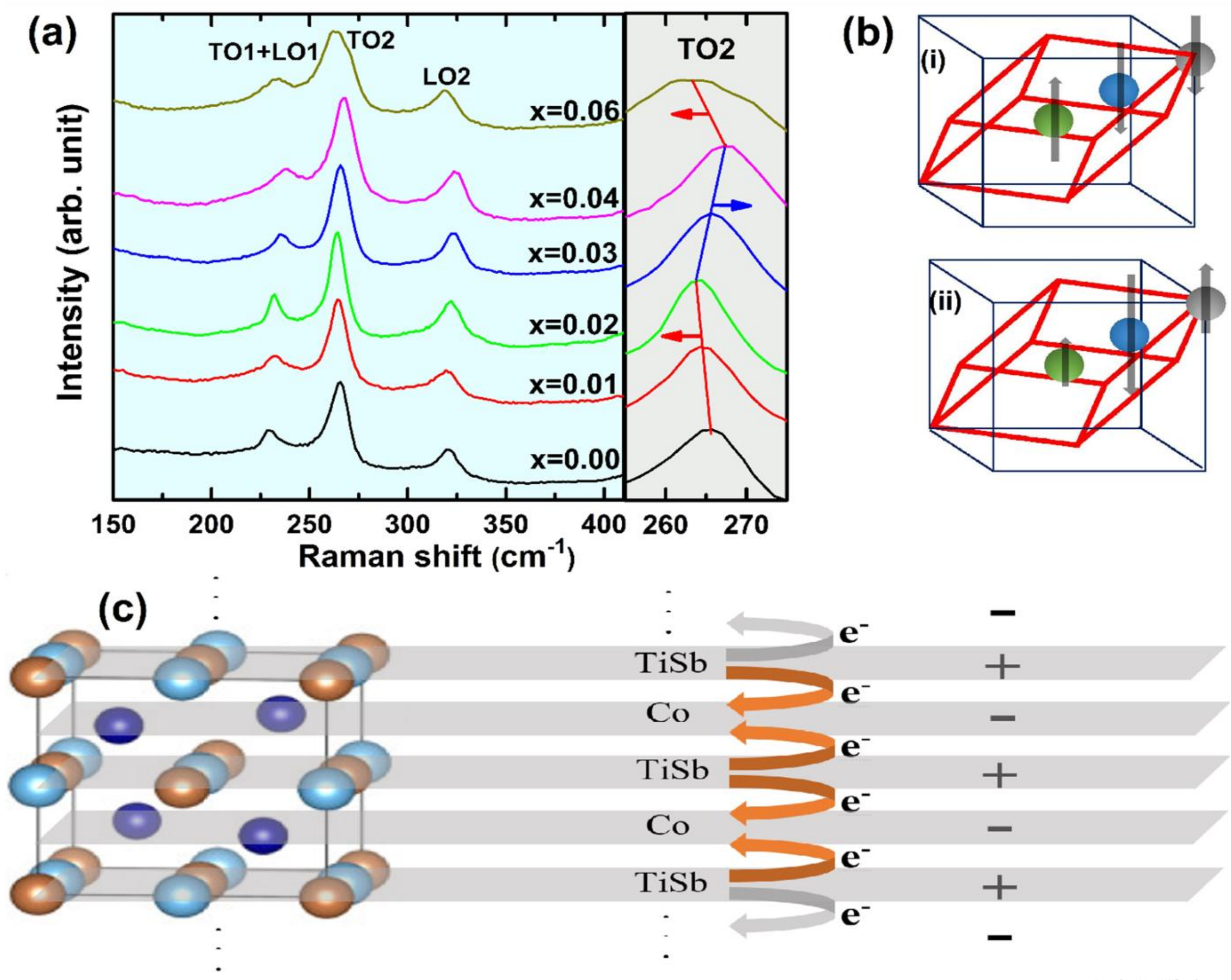

**Fig. 10**. (Color online). (a) Raman spectroscopy data with indexing of vibrational modes for $TiCoSb_{1+x}$ synthesized alloys at room temperature. Inset represents the shift of the transverse optical-2 (TO2) mode. (b) Schematic diagram of vibration for Ti, Co, and Sb atoms in (i) LO1, TO1, (ii) and LO2, TO2 vibrational modes within TiCoSb sample. The arrow is nearly propositional to the atomic displacement during vibration. The green, blue, and grey balls represent Sb, Co and Ti atoms, respectively. (c) Representation of the origin of polar nature in TiCoSb material. The arrangement of $(TiSb)^{+1}$and $Co^{-1}$charged planes within the crystal structure results in an electric dipole moment.

synthesized samples may be recognized as a combination of TO1 and LO1 modes. The other two peaks in fig. 10(a) at 264 $cm^{-1}$ and 318 $cm^{-1}$ may be identified as TO2 and LO2 modes, respectively. Schematic diagram of lattice vibration for LO1, TO1 and LO2, TO2 modes in TiCoSb sample is depicted in fig. 10(b). It is crucial to note that TO2 and LO2 modes in the RS split for the synthesized $TiCoSb_{1+x}$ samples [fig. 10(a)]. Splitting between TO and LO modes arises due to the polar nature of TiCoSb HH alloys. The oxidation states of Ti, Co, and Sb in TiCoSb are +4, -1, and -3, respectively.[8] Hence, TiCoSb structure may be considered as the alternating arrangement of negatively charged $Co^{-1}$ planes and positively charged $(TiSb)^{+1}$ planes [fig. 10(c)].[60] Schematic representation for the charge state of TiSb and Co planes in TiCoSb is shown in fig. 10(c). The alternating $Co^{-1}$ and $(TiSb)^{+1}$ planes create a non-uniform charge distribution along a specific crystallographic direction and lead to polarization in TiCoSb [fig. 10(c)].[60] The observed frequency shift between LO2 and TO2 modes in fig. 10(a) indicates that the synthesized $TiCoSb_{1+x}$ samples are polar in nature. LO phonon vibration is strongly affected by the long-range electric field due to ionic polarization,[61] leads to a frequency shift between LO and TO modes in TiCoSb. Born and Huang have explained the origin of LO-TO splitting and the characteristics of phonons in a crystal using a theoretical model, considering interactions between atoms, including charge carriers and

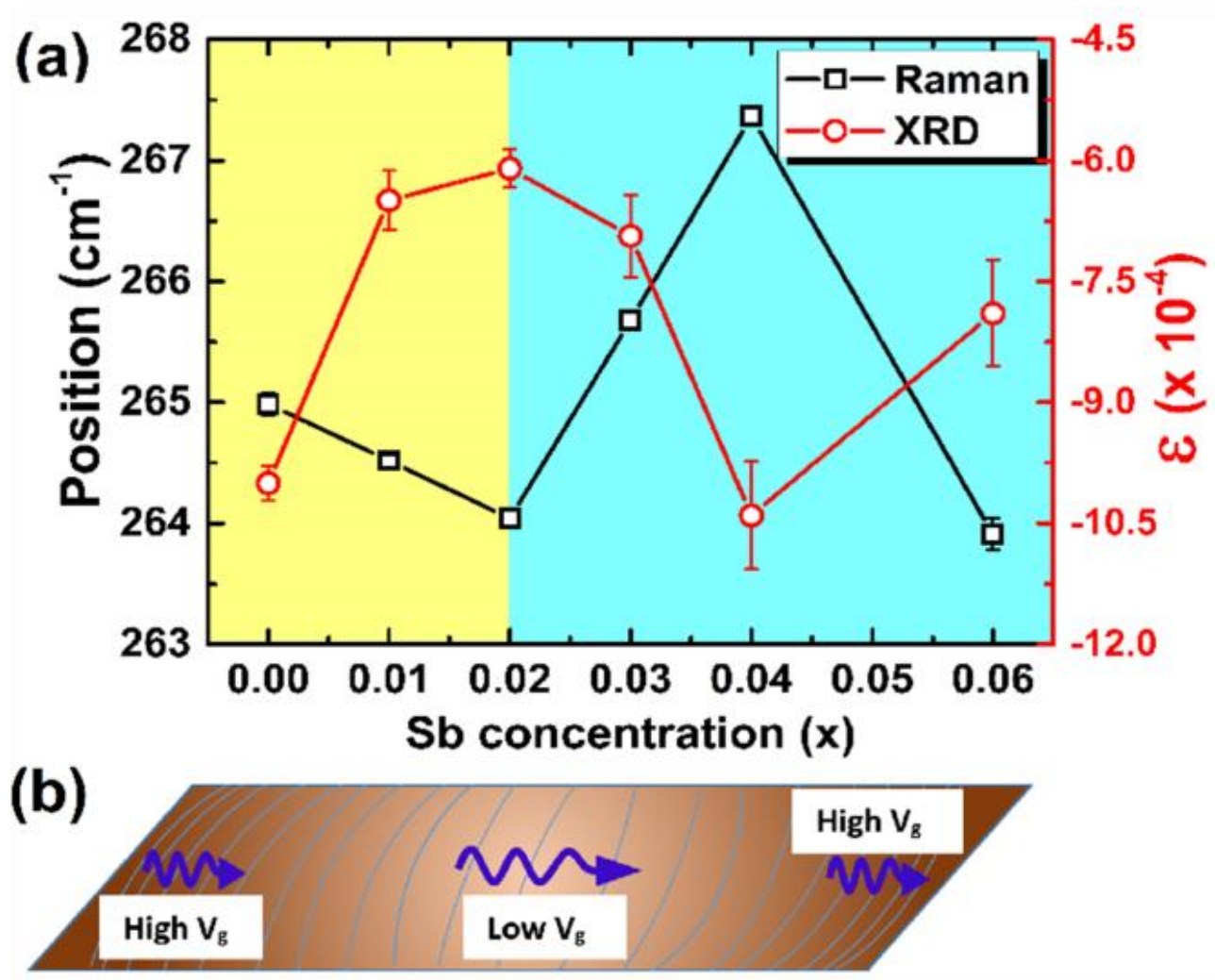

**Fig. 11.** (Color online). (a) Variation in position of the highest intense peak for Raman spectroscopy data and estimated strain (ε) from X-ray diffraction, as a function of Sb concentration for the synthesized $TiCoSb_{1+x}$ HH polycrystalline samples. (b) The schematic represents the effect of strain on group velocity ($v_g$) in the crystal lattice.

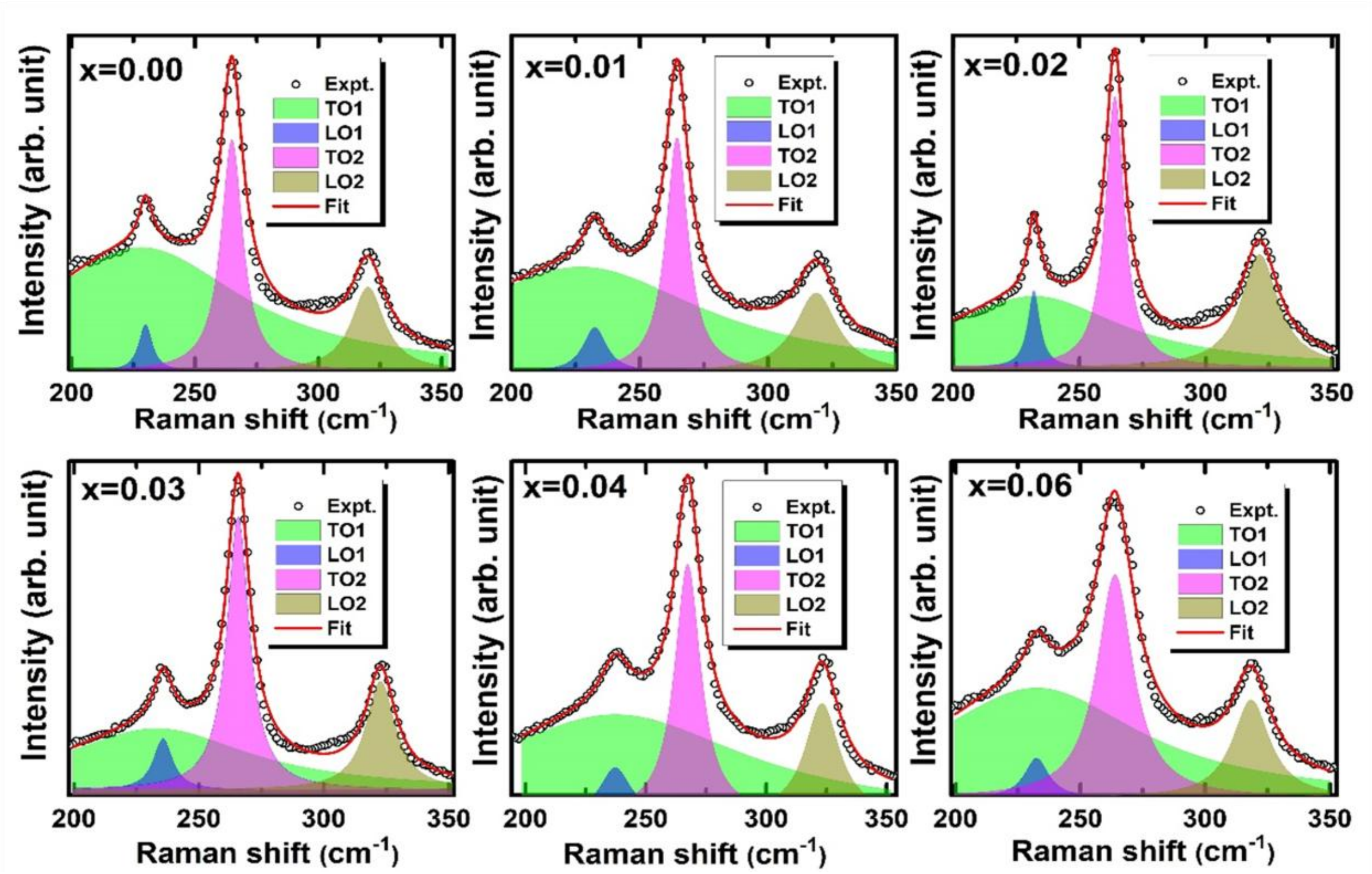

**Fig. 12.** (Color online). Vibrational modes in Raman spectroscopy data of $TiCoSb_{1+x}$ (x=0.0, 0.01, 0.02, 0.03, 0.04, and 0.06). Peaks are fitted with a multiple-peak fitting functions. LO1, TO1, LO2, and TO2 phonon modes are fitted using a Lorentzian function.

vibration.[62] According to the theoretical model, the relation between LO and TO modes is described as,[63]

$$\omega_{LO}{}^2 = \omega_{TO}{}^2 + \Psi(q)\frac{|q|^2}{\Omega}\left(\sum_a \frac{e_q.Z_a.e_{LO}{}^a}{\sqrt{M_a}}\right)^2 \qquad (2)$$

where, $\omega_{LO}$ and $\omega_{TO}$ represent the frequencies of the LO and TO phonon modes, respectively. $\mathbf{e_q}$ represents a unit vector along the momentum direction of q, and $e_{LO}{}^a$ signifies the eigenvector of the dynamical matrix of the LO mode at $|q|\rightarrow 0$. $M_a$ is the mass of atom $a$ in the unit cell, and $Z_a$ represents the Born effective charge of atom $a$. The Born effective charge characterizes the change in polarization in response to the displacement of the atom owing to lattice vibration. Additionally, Ψ(q) represents the Coulomb interaction in Fourier space. Crucial to note that the distribution of charges in a polar material varies during lattice vibration. Displacement of atoms due to lattice vibration causes variation in charge distribution and shifts the overall electric polarization. Hence, $Z_a$ is nonzero for polar material, and LO-TO splitting is observed according to Eq. 2. Co atoms are located at the center of both Ti and Sb tetrahedral in TiCoSb HH alloy.[8] Figure 10($b_{ii}$) correspond to the vibration of Co atoms in the opposite direction relative to Ti and Sb atoms, produce high-energy optical phonon branch owing to the involvement of stronger chemical bonds relative to the vibration as depicted in fig. 10($b_i$) (vibration of Co and Ti atoms in the same direction).[59] However, a substantial $Z_a$ along with vibration of $(TiSb)^{+1}$ and $Co^{-1}$ planes in opposite directions, concomitantly leads to a large LO2-TO2 splitting in TiCoSb alloys.

The position of TO2 mode and strain, obtained from XRD data[32] of the synthesized samples, are depicted in fig. 11. TO2 mode of $TiCoSb_{1+x}$ undergoes red shift for $0 \leq x \leq 0.02$ and blue shift for $0.02 < x \leq 0.04$. Blue and red shifts in RS data may be directly correlated with the variation of compressive strain within the material.[64] Embedded phases and Co vacancy causes modification in strain of synthesized TiCoSb HH alloy (fig. 11).[32] However, a change in compressive strain may locally alter the phonon group velocity due to the development of lattice stiffness.[13] Lattice softening and stiffness are developed for $0 \leq x \leq 0.02$ and $0.02 \leq x \leq 0.04$, respectively, in the synthesized $TiCoSb_{1+x}$ HH alloy due to modification in strain (fig. 11). $\kappa_l$ in material may be defined as [65]

$$\kappa_l = \frac{1}{3}\int_0^{\omega_{max}} v_g^2\, C_s \tau\, d\omega \qquad (3)$$

where $C_s$ is the heat capacity, $v_g$ is the group velocity of phonon through the matrix, and τ is the phonon relaxation time. Equation 3 may be written as, considering phonon–phonon scattering dominates, [66, 67]

$$\kappa_l = \frac{(6\pi^2)^{2/3}\bar{M}}{V^{2/3}4\pi^2\gamma^2}\frac{\langle v_g^3\rangle}{T} = A\frac{\langle v_g^3\rangle}{T} \qquad (4)$$

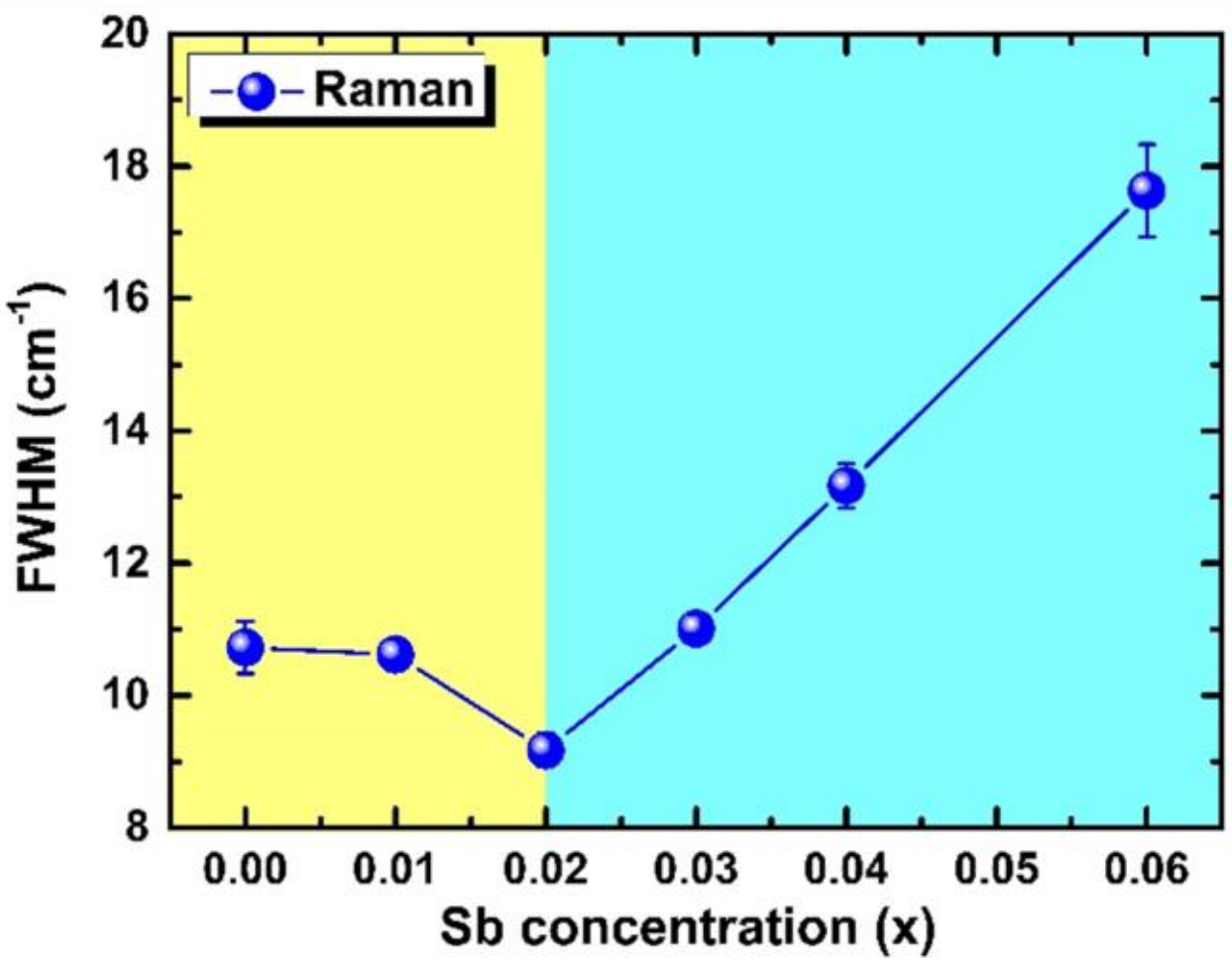

**Fig. 13.** (Color online). Full Width at Half Maxima (FWHM) as a function of Sb concentration, obtained by fitting the most intense peak in Raman spectroscopy, for synthesized $TiCoSb_{1+x}$ ($0.0 \leq x \leq 0.06$) HH alloys.

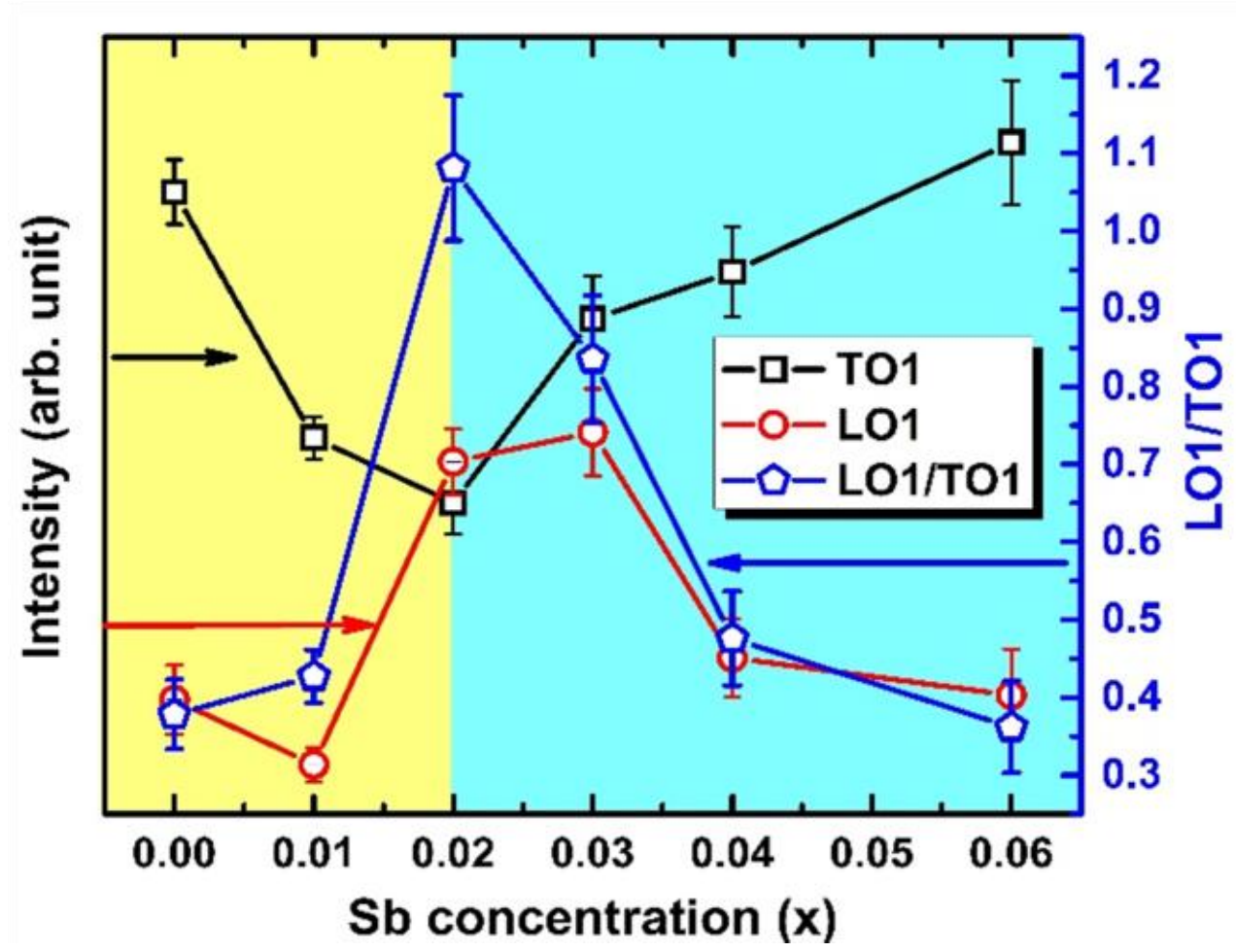

**Fig. 14.** (Color online). Variation in intensity of TO1 and LO1 mode of Raman spectroscopy with Sb concentration for synthesized $TiCoSb_{1+x}$ (x=0.0, 0.01, 0.02, 0.03, 0.04, and 0.06) HH alloys. LO1/TO1 define the intensity of LO1 mode relative to TO1 mode.

where $\langle v_g^3\rangle$ is cubic group velocity of phonon, average over the Brillouin Zone. $\bar{M}$, V and $\gamma$ are atomic mass, unit cell volume and Grüneisen parameter, respectively. Relation between $v_g$ of phonon and $\kappa_l$ is defined through Eq. 4. Further, $v_g$ may be tuned by lattice softening.[13] Lattice softening in synthesized $TiCoSb_{1+x}$ samples is tuned by the lattice strain, which leads to a modification in $\kappa_l$. It is important to note that lattice softening for TE materials has a more significant impact on $\kappa_l$ than micro/nanostructural scattering.[66, 67] Enhancement of TE efficiency by lattice softening, results from alloying or the introduction of vacancies, is one of the crucial and well-known approaches.[14, 15] Hanus et al. have reported that lattice softening causes a modification in $v_g$ from 5830 m/s to 4440 m/s, leads to a roughly 50% decrease in κ for polycrystalline Si.[13] Further, a 7% reduction in $v_g$ by tuning internal strain, results a reduction in κ ~21%, is reported for PbTe.[13]

In order to study the effect of Sb concentration on the crystalline order of $TiCoSb_{1+x}$ samples, the RS data is fitted with a multiple peak function. The fittings are shown in fig. 12. FWHM of the most intense peak (i.e., TO2 mode) behaves anomalously with Sb concentration as presented in fig. 13. FWHM decreases up to x=0.02 and thereafter increases, indicates that most ordered sample is $TiCoSb_{1.02}$. It also supports the analysis of XRD and XANES data. Further, the peak around 230 cm$^{-1}$ is fitted with two Lorentzian type functions to accommodate the TO1 and LO1 modes. A non-monotonic variation in intensity for LO1 and TO1 modes, as deciphered from RS data, is depicted in fig. 14. Figure 12 and fig. 14 indicate that TO1 mode is dominant for either side of x=0.02, whereas LO1 mode is prevailed over for x=0.02 in synthesized samples. However, both the LO1 and TO1 modes equally contribute in RS for $TiCoSb_{1.02}$ sample i.e., the relative intensity $\left(\frac{LO1}{TO1}\right)$ is maximum (~ 1), as shown in fig. 14. It is crucial to note that the lattice softening effect due to modification in compressive strain is corroborated with the relative contribution of LO1/TO1 at 230 cm$^{-1}$ peak of RS for $TiCoSb_{1+x}$ synthesized samples.

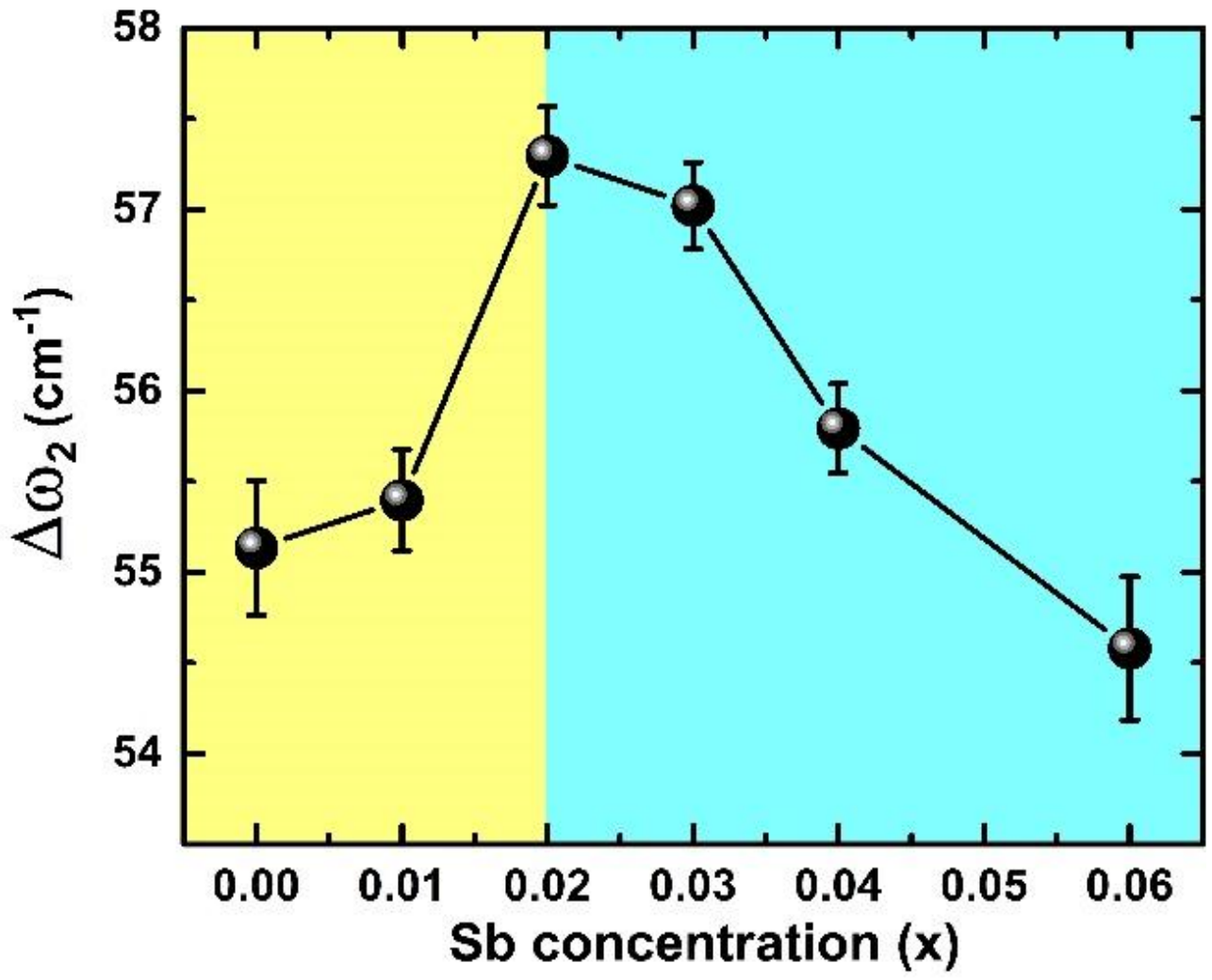

**Fig. 15.** (Color online). The variation of the frequency difference, $\Delta\omega_2$ ($\Delta\omega_2$ =LO2-TO2) as a function of Sb concentration for synthesized $TiCoSb_{1+x}$ HH alloy.

Frequency difference between the TO2 and LO2 modes, $\Delta\omega_2 = \omega_{LO2} - \omega_{TO2}$ as a function of Sb concentration for $TiCoSb_{1+x}$ synthesized samples, is depicted in fig. 15. $\Delta\omega_2$ is maximum for the $TiCoSb_{1.02}$ sample and reduces on either side. Variation of $\Delta\omega_2$ with Sb concentration is corroborated with change in strain and Co vacancy in $TiCoSb_{1+x}$ samples (fig. 15). It is crucial to note that maximum crystalline order along with host phase is reported for $TiCoSb_{1.02}$.[32] $\Delta\omega_2$ increases for 0.0≤x≤0.2 due to rise in scattering phase space.[68] Minimum compressive strain and presence of Co vacancy lead to high scattering phase space.[68] However, the decrease of $\Delta\omega_2$ for $x > 0.02$ seems to be related to the screening effect due to the large amount of Co vacancy.[61] The presence of free carriers and concomitant screening effect may modify both the electron-LO phonon interaction and the dispersion characteristic of the LO-optical phonon. According to the Thomas-Fermi approximation, the Coulomb interaction in phase space (q-space) in the presence of screening effect may be written as [61, 63]

$$\Psi(q) = \frac{e^2 r_{screening}^2}{\varepsilon_\infty (1 + |q|^2 r_{screening}^2)} \quad (5)$$

where $r_{screening}^2$ is the Thomas-Fermi screening length (or Debye length), and is defined as, [61]

$$r_{screening}^2 = \frac{ek_BT}{ne^2} \quad (6)$$

where, $k_B$ is the Boltzmann constant, and n represents carrier concentration. Equation 6 indicates that higher n leads to a better screening effect and a decrease in $\Psi(q)$. Equation 2 indicates that the value of $\Psi(q)$ is directly related to splitting of vibration modes in RS spectra. Crucial to note that value of n increases in $TiCoSb_{1+x}$ for x>0.02 due to Co vacancy, enhances the screening effect.[32] Increase in Co vacancy is also supported by the EXAFS study in this article. The heightened screening in $TiCoSb_{1+x}$ (x>0.02) leads to a reduction in LO2-TO2 splitting and concomitantly decreases polar optical phonon scattering.

## IV. Conclusion

In this work, efforts are employed to reveal the details of lattice vibration and concomitantly, the effects of defects and crystalline strain on the κ of $TiCoSb_{1+x}$ (x=0.0, 0.01, 0.02, 0.03, 0.04, and 0.06) synthesized materials. Room temperature κ of synthesized HH alloys is obtained from XRD data as well as RS data, and temperature-dependent κ is calculated from the thermal variation of RS data, employing the optothermal Raman technique. It is noteworthy to mention that the nature of lattice vibration is explored through RS measurements, and vibrational modes are reported for the TiCoSb HH alloy. The shiny phases in SEM images and analysis of EDS data confirm the presence of metallic embedded phases along with Co vacancies. HRTEM data accompanied with SAED patterns of the synthesized alloys show evidence of synthesized polycrystalline TiCoSb HH alloys and contraction in lattice volume with increasing Sb concentration. The systematic decrease in d-spacing for (h k l) planes may be related to Co vacancies as explored in Rietveld refinement, SEM, and EDS data. The Co vacancies lead to a change in oxidation states of Co atoms in the synthesized $TiCoSb_{1+x}$ alloys. XANES measurements are performed to obtain experimental evidence for the formation of Co vacancies along with embedded phases in $TiCoSb_{1+x}$ synthesized polycrystalline HH alloy. The detailed structural analysis reveals that Sb concentration strongly influences the structural parameters and embedded phases in the TiCoSb host matrix. It is important to mention that maximum host phases in a company with minimum Co vacancies and structural distortion are obtained for $TiCoSb_{1.02}$ (x=0.02) sample. However, it is crucial to mention that the arrangement of nearest neighbour and average coordination number are probed through the EXAFS experimental facility, primarily for the $TiCoSb_{1+x}$ synthesized alloy. XANES data of Co K edge spectra also reveal the Co vacancy in synthesized samples, and the most ordered phase is $TiCoSb_{1.02}$. However, the presence of Co vacancies is strongly regulated by tuning of Sb concentration in the TiCoSb HH alloy. EXAFS data reveals that a low value of $\sigma^2$ and an insignificant change in volume of the 2nd shell indicate that the crystalline order of TiCoSb HH alloy in the short range is mainly preserved due to the 2nd shell i.e., Co-Sb. The insight into the bonding states of the constituent elements in the synthesized HH alloys is explored by XPS. The charge transfer from Ti and Co atoms to Sb atoms confirms the formation of HH alloys as indicated by XRD, EDS, and HRTEM data. Deconvolution of the Co core spectrum of XPS data reveals that the area under the curve of the peak, which arises in higher oxidation states, increases with Sb concentration, indicates that an elevated amount of Co possesses a higher oxidation state. The outcome is well aligned with the XANES result. The effect of Sb concentration on phonon scattering for $TiCoSb_{1+x}$ synthesized samples is revealed by RS. $\varepsilon$, estimated using XRD, and the positions of the most intense peak in RS data are nicely corroborated. Minimum compressive strain is obtained for $TiCoSb_{1.02}$. Lattice stiffness due to internal compressive strain in $TiCoSb_{1+x}$ leads to a local change in phonon frequencies in the synthesized alloy and modifies the group velocity ($\upsilon_g$) of the phonon, results an alteration in $\kappa_l$. Minimum $\upsilon_g$ corresponding to the lowest $\kappa$ is reported for the x=0.02 sample. Further, a gradual decrease in compressive strain for $0.0 \leq x \leq 0.02$ induces scattering phase space and concomitantly decreases $\kappa_l$. However, for $0.02 < x \leq 0.06$, $\kappa_l$ increases owing to an increase in compressive strain and Co vacancies. LO-TO splitting and the polarization field are reduced for $0.02 < x \leq 0.06$ due to the screening effect, results from an increase in Co vacancies. It is noteworthy to mention that, $\kappa_l$ is lowest for $TiCoSb_{1.02}$ i.e., sample with maximum host phase (TiCoSb) amid the synthesized $TiCoSb_{1+x}$ HH polycrystalline alloy. Lattice softening and LO-TO splitting play a crucial role in thermal energy transport through lattice vibration, and $\kappa_l$ may be tuned through crystalline strain and Co vacancies in TiCoSb HH alloy.

**Supporting Information**

Supporting Information accommodates the detailed calculation of thermal conductivity, figures of temperature-dependent Raman spectroscopy, figures of energy dispersive X-ray analysis, results obtained from analysis of selected area electron diffraction pattern in TEM data, working principals of the instrument and procedure of analysis for X-ray absorption fine structure data, and fitted curves of X-ray photoelectron spectroscopy (XPS).

## Acknowledgments

This work is supported by the Science and Engineering Research Board (SERB), Government of India, in the form of a sanctioned research project (File Number: EEQ/2018/001224). The financial support of UGC-DAE-CSR Kalpakkam, India (Ref: CRS/2021-22/04/639, CRS/2022-23/04/893) and UGC-DAE-CSR Mumbai, India (Ref: CRS/21-22/03/365) is acknowledged in the form of research project grants. We are thankful to Dr. Aritra Banerjee for the discussion and help in carrying out Raman spectroscopic measurements. We are also thankful to Dr. Shubhankar Roy, Vidyasagar Metropolitan College, for partial assistance in synthesizing samples. Author SM is thankful to CSIR, India, for providing Research Fellowships. Authors would like to acknowledge DST-PURSE, Phase-II, for the procurement of Raman spectroscopy at the Department of Physics, CU. We also thank Prof. Ram Janay Choudhary, UGC-DAE CSR, Indore, for his assistance with XPS measurement. Further, we would also like to acknowledge the synchrotron beamline, BL-9, INDUS-2, Raja Ramanna Centre for Advanced Technology, for providing the facility of XAFS measurement.

---

*Corresponding Author
Email: kartick.phy09@gmail.com

# Supporting Information

## Drastic modification in thermal conductivity of TiCoSb Half-Heusler alloy: Phonon engineering by lattice softening and ionic polarization

Suman Mahakal[1], Avijit Jana[1], Diptasikha Das[2], Nabakumar Rana[3], Pallabi Sardar[2], Shamima Hussain[4], Santanu K. Maiti[5], and Kartick Malik*[1]

[1] *Department of Physics, Vidyasagar Metropolitan College, Kolkata-700 006, India.*
[2]*Department of Physics, Adamas University, Kolkata-700 126, India.*
[3]*Department of Physics, University of Calcutta, Kolkata-700 009, India.*
[4]*UGC-DAE Consortium for Scientific Research, Kalpakkam Node, Kokilamedu-603 104, Tamil Nadu, India.*
[5]*Physics and Applied Mathematics Unit, Indian Statistical Institute, 203 Barrackpore Trunk Road, Kolkata-700 108, India.*

*Corresponding author:* kartick.phy09@gmail.com

## Table of contents:

## Section 1: Working principle of the instrument for X-ray Absorption Fine Structure (XAFS) spectroscopy

XAFS data of $TiCoSb_{1+x}$ (x=0, 0..01, 0..02, 0.03, 0.04, and 0.06) samples were carried out at Co K-edge (7708.9 eV) in transmission mode. The beamline operates in the energy range of 4 keV to 25 keV with energy resolution E/dE∼ 10. The beamline optics consist of a Rh/Pt coated collimating meridional cylindrical mirror followed by a Si (111) (2d = 6.2709 Å) based double crystal monochromator (DCM). Horizontal focusing was achieved using the second crystal of DCM, a sagittal cylindrical crystal. One ionization chamber with a length of 300 mm and a solid-state detector were used in the setup. The photon energy scale was calibrated by measuring the XAFS spectrum of a standard Co metal foil. The absorption coefficient μ was obtained using the relation $\mu(E) \propto I_0/I_t$, where $I_0$ is the incident X-ray intensity, and $I_t$ is the measured transmission intensity.

## Section 2: Details of thermal conductivity calculation

In order calculate total thermal conductivity κ ($\kappa=\kappa_e + \kappa_L$, $\kappa_e$ = electronic thermal conductivity and $\kappa_L$= lattice thermal conductivity), we have estimated $\kappa_e$ and $\kappa_l$ seperately. $\kappa_e$ is estimated employing Wiedemann Franz law, $\kappa_e = L\sigma T$ (L is the Lorenz number). Further, L is evaluated from the temperature dependent Seebeck coefficient [*S(T)*] data, using the relation,[1] $L = [1.5 + exp(-|S|/116)] \times 10^{-8}$. The $\sigma$ and *S(T)* data are presented elsewhere.[2] Further $\kappa_L$ is estimated using Slack equation. $\kappa_L$,within the limit Umklapp scattering processes may be written as,[3]

$$\kappa_L = \Lambda \frac{\bar{M}\theta_D^3\delta}{\gamma_G^2 n^{2/3} T} \tag{S1}$$

where $\bar{M}$, $\delta^3$, and n are average atomic mass in the unit cell, volume per atom and number of atoms in the primitive cell, respectively. Details can be found elsewhere for $TiCoSb_{1+x}$ synthesized samples.[2]

## Section 3: Temperature-dependent Raman spectroscopy

**Figure S1:** (Color online). Temperature-dependent Raman spectroscopy of $TiCoSb_{1+x}$ (x=0, 0.02, and 0.04) HH polycrystalline alloy.

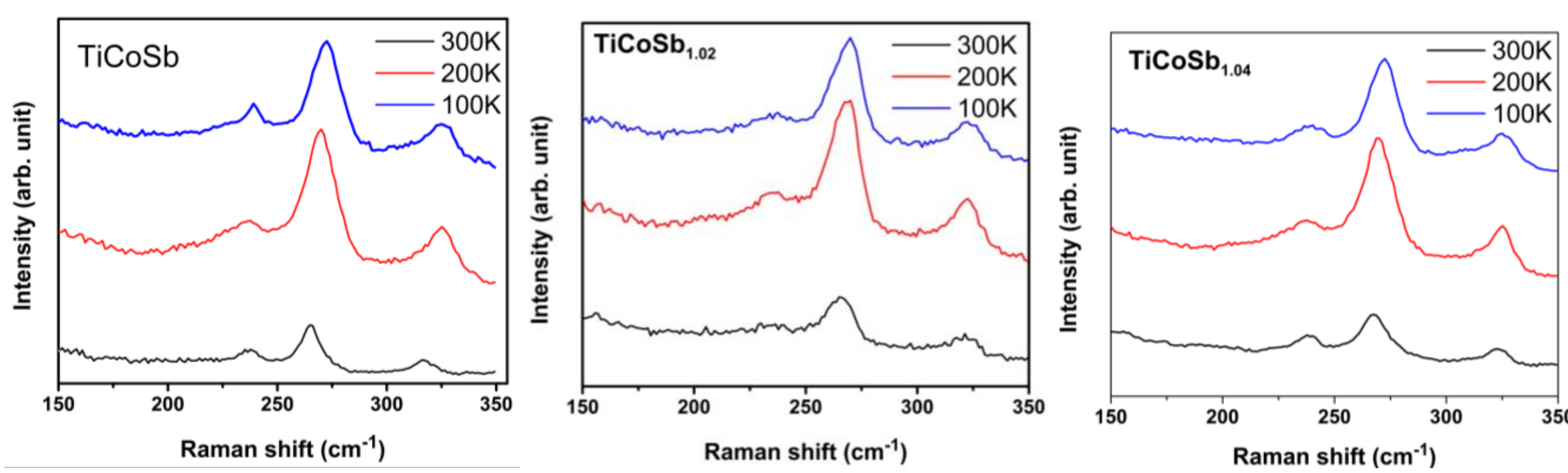

**Figure S2:** (Color online). Energy Dispersive X-ray (EDX) spectroscopy of the synthesized $TiCoSb_{1+x}$ (x = 0.00, 0.01, 0.02, 0.03, 0.04, and 0.06) Half-Heusler alloys at room temperature.

**Sum Spectrum**

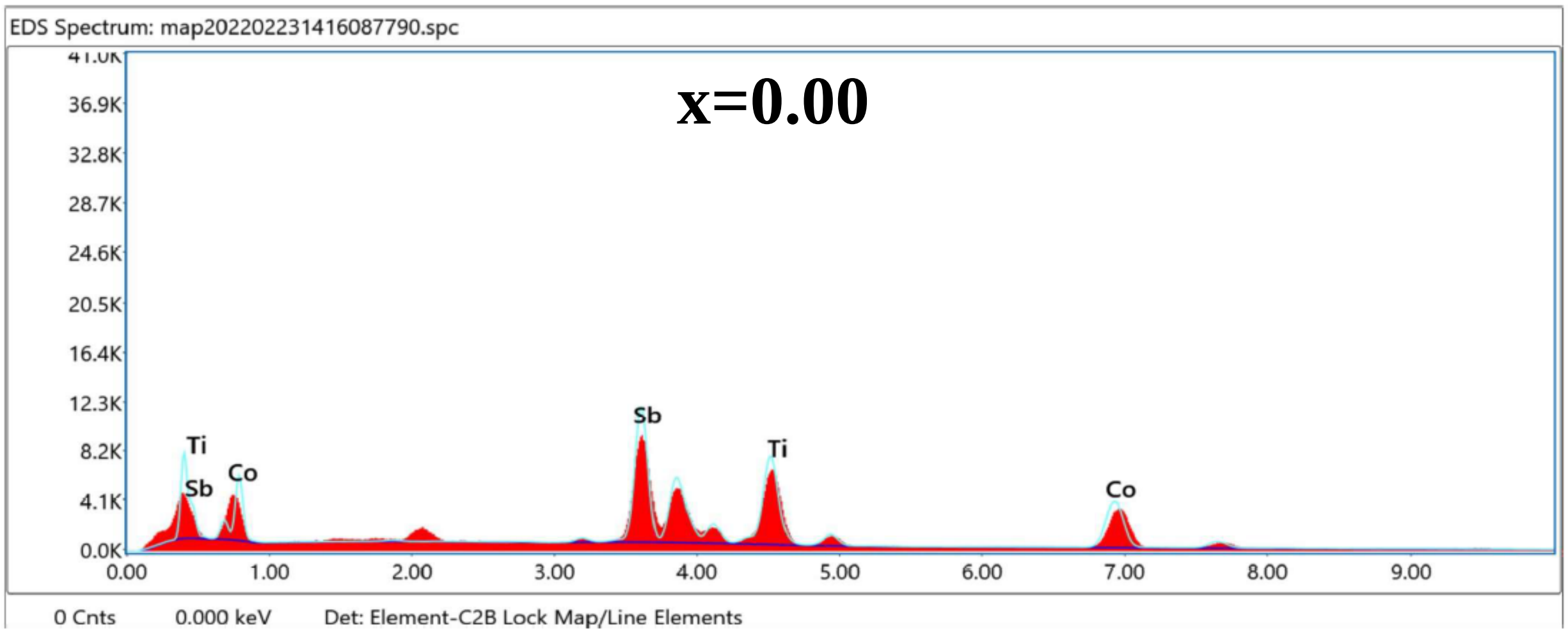

**Smart Quant Results**

| Element | Weight % | Atomic % | Net Int. | Error % |
|---|---|---|---|---|
| SbL | 52.53 | 32.68 | 251.61 | 2.33 |
| TiK | 21.26 | 33.64 | 176.38 | 3.64 |
| CoK | 26.21 | 33.68 | 113.08 | 3.76 |

**Sum Spectrum**

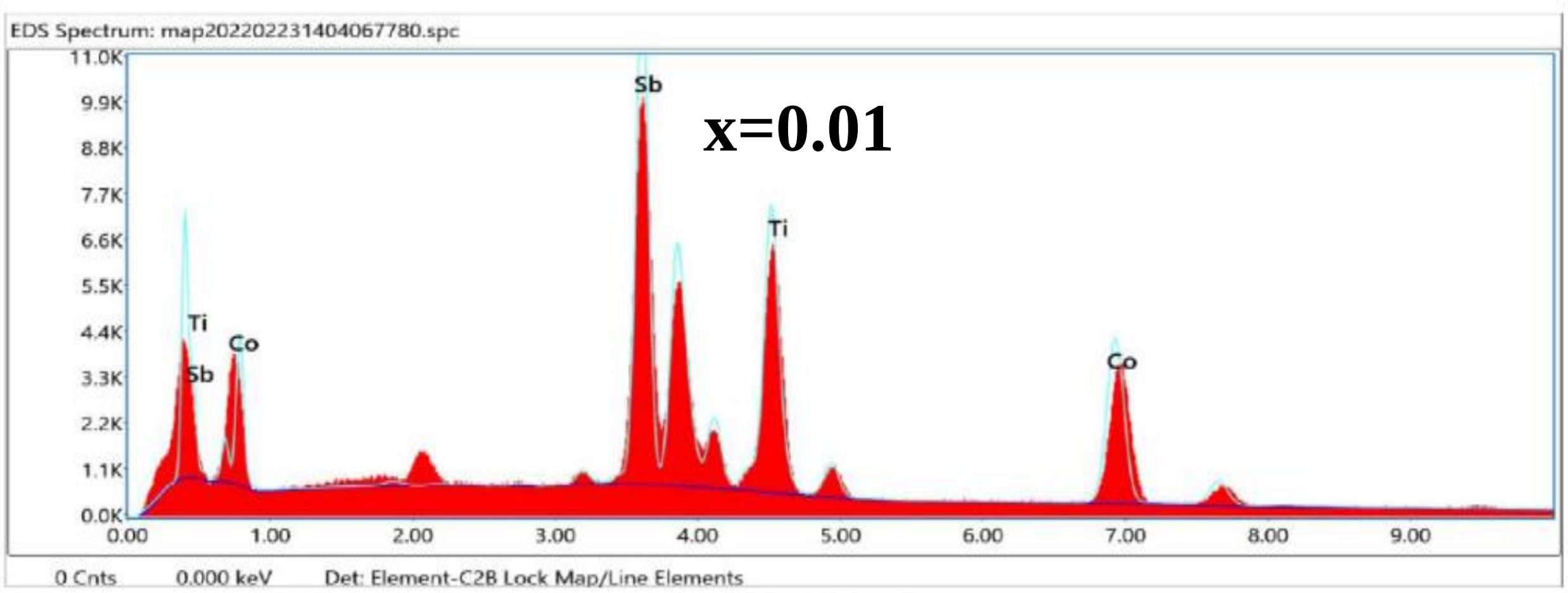

**Smart Quant Results**

| Element | Weight % | Atomic % | Net Int. | Error % |
|---|---|---|---|---|
| SbL | 52.99 | 33.07 | 269.73 | 2.28 |
| TiK | 21.13 | 33.55 | 166.38 | 3.73 |
| CoK | 25.88 | 33.38 | 118.09 | 3.73 |

**Sum Spectrum**

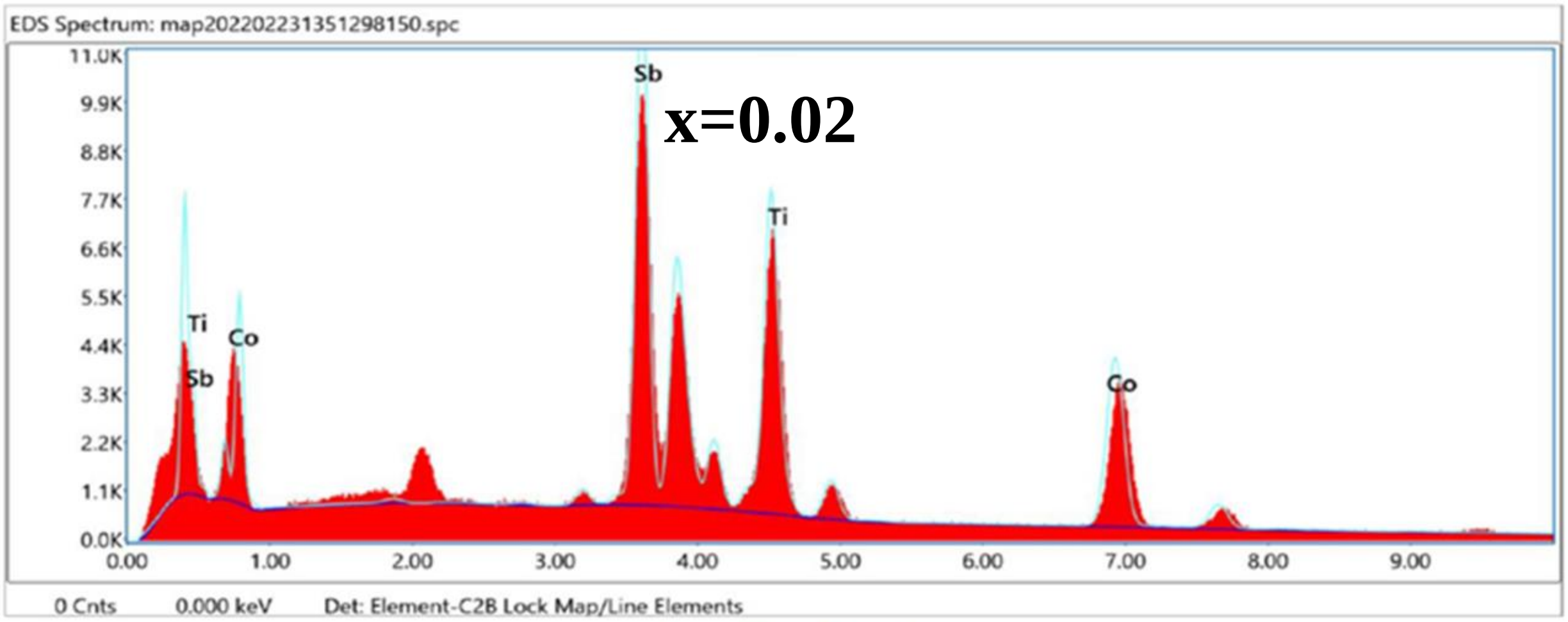

**Smart Quant Results**

| Element | Weight % | Atomic % | Net Int. | Error % |
|---|---|---|---|---|
| SbL | 53.14 | 33.21 | 267.04 | 2.31 |
| TiK | 21.09 | 33.52 | 176.82 | 3.70 |
| CoK | 25.77 | 33.27 | 114.01 | 3.77 |

**Sum Spectrum**

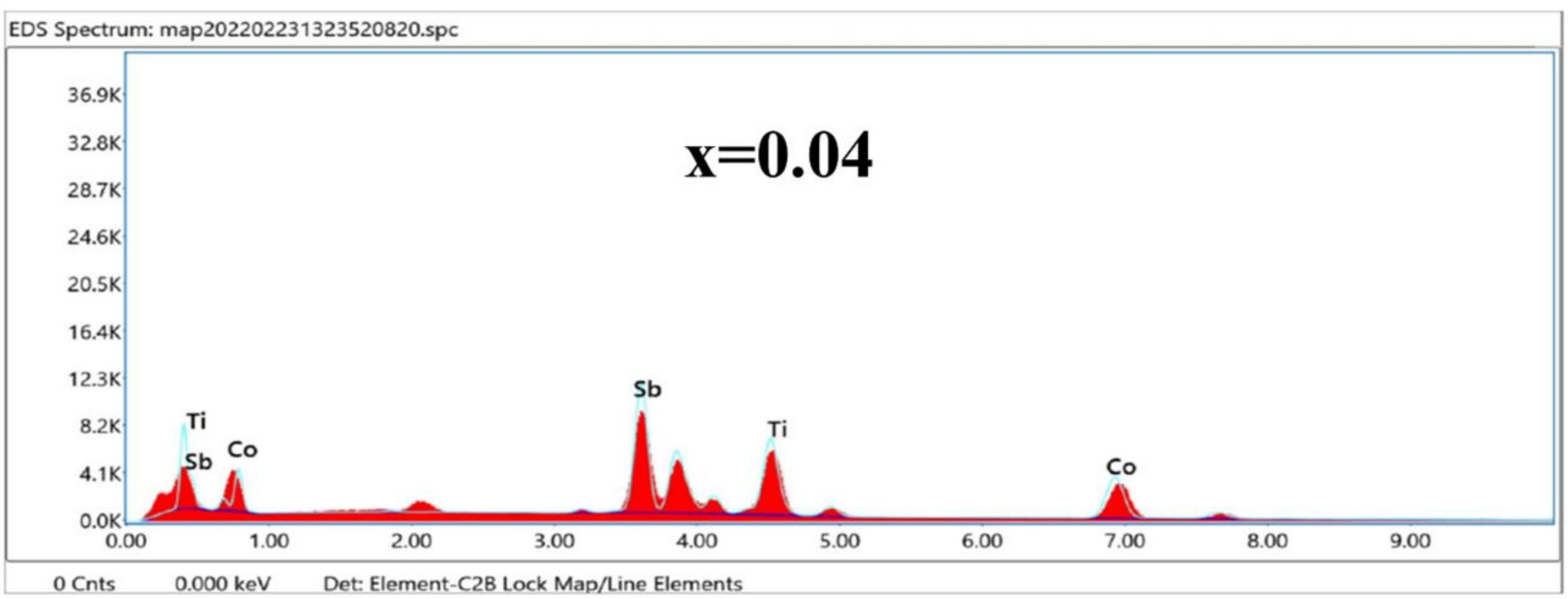

**Smart Quant Results**

| Element | Weight % | Atomic % | Net Int. | Error % |
|---|---|---|---|---|
| SbL | 53.75 | 33.76 | 238.17 | 2.32 |
| TiK | 20.76 | 33.16 | 151.67 | 3.75 |
| CoK | 25.49 | 33.08 | 99.37 | 3.80 |

**Sum Spectrum**

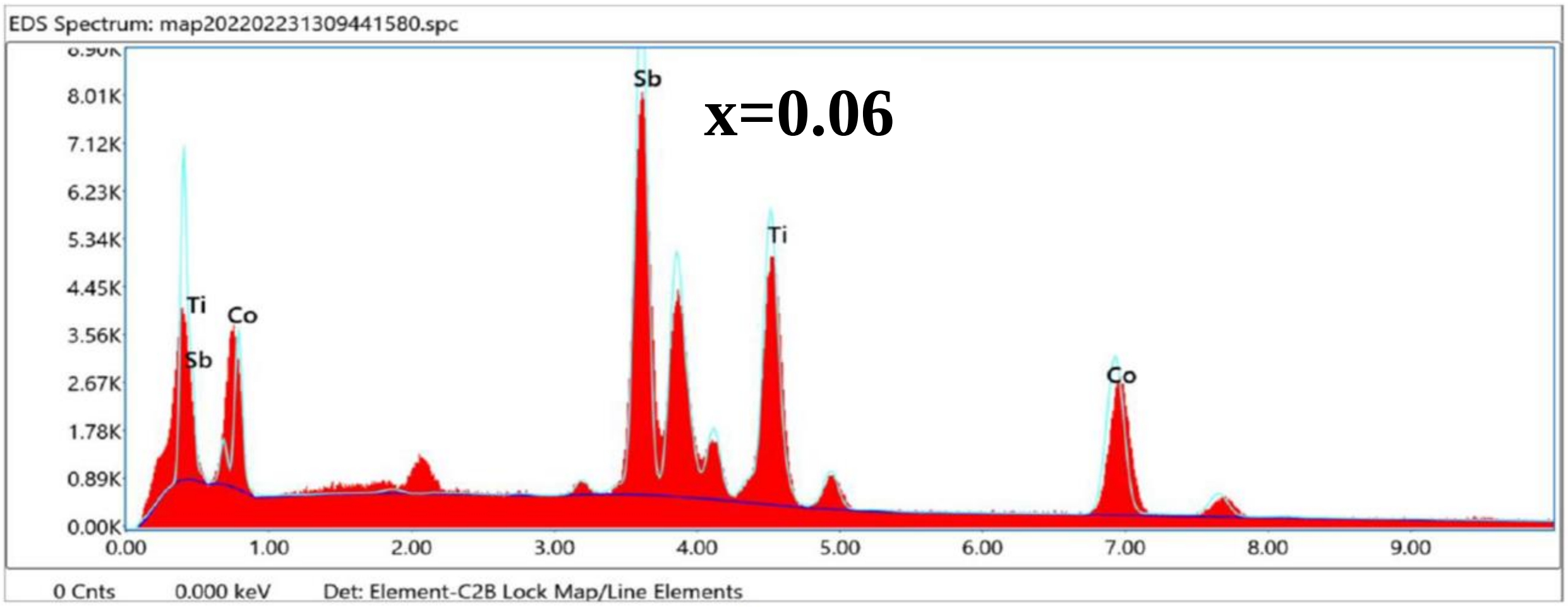

**Smart Quant Results**

| Element | Weight % | Atomic % | Net Int. | Error % |
|---|---|---|---|---|
| SbL | 53.39 | 33.50 | 197.76 | 2.33 |
| TiK | 20.25 | 32.33 | 122.55 | 3.80 |
| CoK | 26.36 | 34.17 | 81.49 | 3.89 |

**EDX spectroscopy data analysis and estimation of embedded phases**

1. **EDX spectroscopy data for x=0.0**

$Ti_{0.3364}Co_{0.3368}Sb_{0.3268}$ => $TiCoSb_{0.97}$ => 97% (TiCoSb) + 3% CoTi

2. **EDX spectroscopy data for x=0.01**

$Ti_{0.3355}Co_{0.3338}Sb_{0.3307}$ => $TiCo_{0.9949}Sb_{0.9856}$ => 98.56% TiCoSb + 1.46% CoTi + 0.5%$V_{Co}$

3. **EDX spectroscopy data for x=0.02**

$Ti_{0.3352}Co_{0.3327}Sb_{0.3321}$ => $TiCo_{0.993}Sb_{0.991}$ => 99.1% TiCoSb + 0.9% CoTi + 0.7%$V_{Co}$

4. **EDX spectroscopy data for x=0.04**

$Ti_{0.3316}Co_{0.3308}Sb_{0.3376}$ => $Ti_{0.982}Co_{0.979}Sb$ => 98.22% TiCoSb + 1.78% CoSb + 2%$V_{Co}$

5. **EDX spectroscopy data for x=0.06**

$Ti_{0.3233}Co_{0.3417}Sb_{0.3350}$ => $Ti_{0.965}Co_{1.02}Sb$ => 96.50% TiCoSb + 3.5% CoSb + 2%Excess-Co

## Section 5: Selected area electron diffraction (SAED) data analysis

**Table S1:** The d-spacing and corresponding (h k l) planes as estimated from the SAED patterns of synthesized $TiCoSb_{1+x}$ (x = 0.00, 0.02, and 0.04) Half-Heusler alloys.

| (h k l) plane | d-spacing (Å) | | | |
|---|---|---|---|---|
| | TiCoSb | $TiCoSb_{1.02}$ | $TiCoSb_{1.04}$ | JCPDS # 300083 |
| 1 1 1 | 3.4637 | 3.4159 | 3.4088 | 3.397 |
| 2 0 0 | 3.0637 | 2.9437 | 2.9362 | 2.942 |
| 2 2 0 | 2.0985 | 2.0729 | 2.0686 | 2.080 |
| 3 1 1 | 1.8282 | 1.8182 | 1.7934 | 1.774 |
| 2 2 2 | 1.7208 | 1.7369 | 1.7248 | 1.698 |
| 4 0 0 | 1.5174 | 1.4704 | 1.4862 | 1.471 |
| 3 3 1 | 1.3775 | 1.3420 | 1.3332 | 1.349 |
| 4 2 2 | 1.2172 | 1.1810 | 1.1951 | 1.132 |
| 5 1 1 | 1.1483 | 1.1178 | 1.1105 | 1.040 |
| 4 4 0 | 1.0514 | 1.0402 | 1.0256 | 0.990 |

## Section 6: X-ray Absorption Fine Structure (XAFS) Spectroscopy data analysis procedure

A promising technique for exploring the atomic environment of a particular element in a material is XAFS spectroscopy. It provides critical insights into structural aspects such as crystalline order, disorder, and atomic defects. XAFS data consists of two distinct regions:

1. X-ray Absorption Near Edge Structure (XANES) – This region primarily reflects electronic properties, including the valence state of the absorbing atom, and is highly sensitive to its local coordination environment.[4]
2. Extended X-ray Absorption Fine Structure (EXAFS) – This portion of the spectrum provides detailed information about the local atomic arrangement, extending over a wider range of neighboring atoms.[5]

Room temperature XAFS data highlighting the XANES and EXAFS part of synthesized samples, obtained in the range 7600 eV to 8500 eV, are shown in fig. S3. In the TiCoSb system, Co has an electronic configuration of [Ar] $3d^9 4s^0$. Two distinct types of electronic transitions contribute to the XAFS spectra. The XANES region in the Co K-edge spectrum arises from the excitation of a 1s electron to the unoccupied 4p states. However, the EXAFS spectrum is obtained due to the excitatuion of an electron from any occupied to the continuum level. The EXAFS spectrum provides valuable information regarding the short-range periodic arrangement of atoms, offering insights into local structural order. In order to analyze the EXAFS data of the synthesized $TiCoSb_{1+x}$ samples, the oscillation part of the measured absorption co-efficient [μ(E)] in XAFS data is converted into the fine structure-function (χ(E)), using the following equation,[6]

$$\chi(E) = \frac{\mu(E) - \mu(E_0)}{\Delta\mu(E_0)} \quad \text{(S2)}$$

where $E_0$ is the threshold energy of transition, $\Delta\mu_0(E_0)$ is edge step and $\mu_0(E)$ is smooth background function as the absorption of an isolated atom. Analysis in k-space is performed by converting the fine structure function of absorption co-efficient χ(E) to k-space χ(k) using the following relation,[6]

$$k = \sqrt{2m_e(E - E_0)\hbar^2} \quad \text{(S3)}$$

where $m_e$ is the mass of the electron, and $\chi(k)$ is weighted by $k^2$ to amplify oscillations at high k. The $k^2\chi(k)$ function is Fourier transformed to real space (R-space) using the ARTEMIS software package.[7] Crystallographic inputs of the TiCoSb phase are accounted for the fitting of the spectra. Fitting in the R space is performed between 1 Å to 4.5 Å. $\chi(R)$ versus R (radial distance) plots are used for further analysis. Multiplication of $k^2$-weighted is performed during Fourier transformation. The quality of the fit is measured by minimizing statistical parameter, R-factor ($R_f$). $R_f$ is defined by the following equation[6, 8]

$$R_f = \frac{\sum_{j=min}^{max}[Re\left(\chi_d(r_j) - \chi_t(r_j)\right)^2 + Im(\chi_d(r_j) - \chi_t(r_j))^2]}{\sum_{j=min}^{max}[Re\left(\chi_d(r_j)\right)^2 + Im\left(\chi_d(r_j)\right)^2]}. \quad \text{(S4)}$$

Four path parameters ($S^2_0$= amplitude reduction factor, $\Delta r$ = change in half path length, $\Delta E_0$ = energy shift, $\sigma^2$= mean square displacement) are taken into consideration to fit the experimental data with a software program, Full Electronic File Format (FEFF) generated modelling using ARTEMIS, which includes IFEFFIT integration.[7] Other multi-scatterings are discared to minimize the number of free variables ($N_{free}$) and to satisfy Nyquist theorem. In order to obtain good fit of the data, multiple scatterings with lower rank than 20.69 is rejected. $N_{free}$ is always kept below the upper limit, set by the Nyquist theorem ($N_{free} \leq N_{Nyquist} = 2\Delta k\Delta r/\pi+1$ ), where $\Delta r$ is the width of the filter windows in the R-space, $\Delta k$ is the actual interval of the fit in the k-space.[9] Here, the total number of neighbors for each shell is kept constant with the intention of minimizing $N_{free}$.

**Figure S3:** (Color online). X-ray absorption spectroscopy at Co K-edge for $TiCoSb_{1+x}$ (x = 0.00, 0.01, 0.02, 0.03, 0.04, and 0.06) synthesized samples at room temperature.

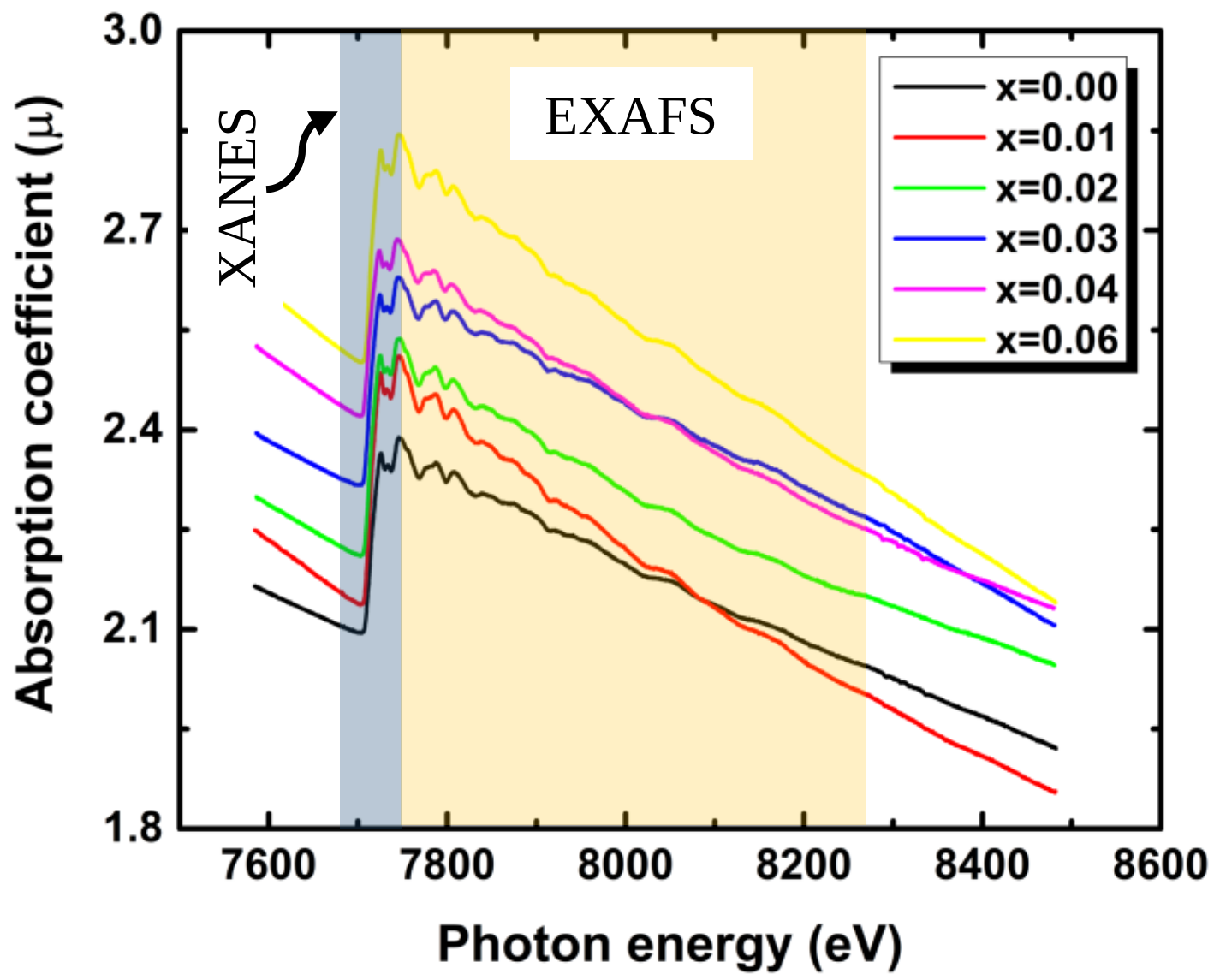

**Section 7: X-ray photoelectron spectroscopy (XPS) data**

**Figure S4:** (Color online). X-ray photoelectron spectroscopy (XPS) analysis of $TiCoSb_{1+x}$ HH polycrystalline materials. Survey spectrum of (a) $TiCoSb_{1.02}$ (b) $TiCoSb_{1.04}$, confirm the formation of $TiCoSb_{1+x}$ alloy. High-resolution XPS spectrum of the Sb 3d core level of (c) $TiCoSb_{1.02}$ (d) $TiCoSb_{1.04}$ HH alloys. High-resolution XPS spectrum of the Ti 2p core level of (e) $TiCoSb_{1.02}$ (f) $TiCoSb_{1.04}$ HH alloys.

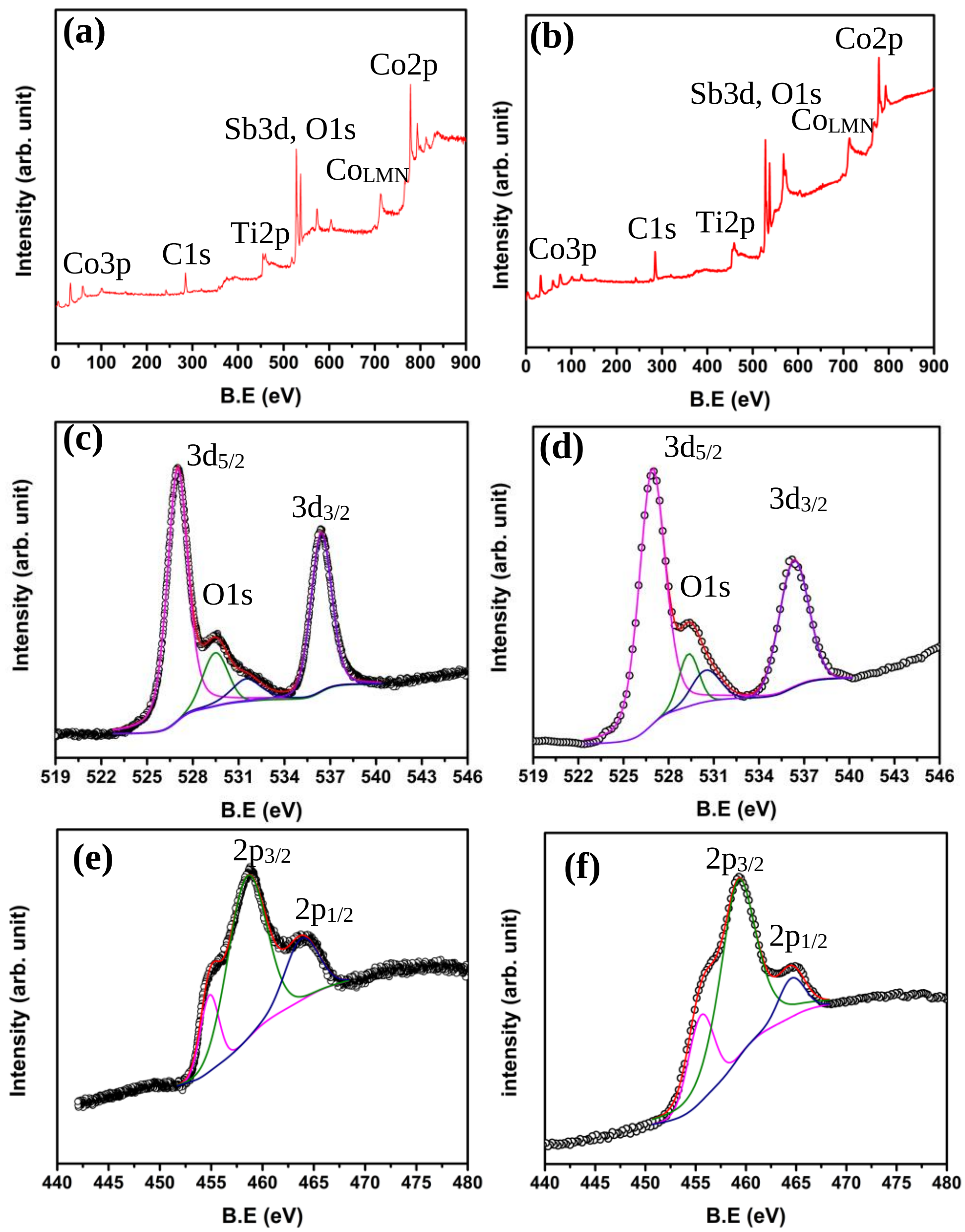

**Section 8: References**